\documentclass[12pt, twocolumn]{aastex631}

\usepackage{amsmath}
\usepackage[utf8]{inputenc}

\UseRawInputEncoding
\usepackage{savesym}
\savesymbol{tablenum}
\usepackage{siunitx}
\restoresymbol{SIX}{tablenum}

\received{XXX}
\revised{YYY}
\accepted{ZZZ}

\submitjournal{APJ}

\shorttitle{Luminosity functions of supernova host galaxies}
\shortauthors{Z. LIANG et al.}

\begin{document}

\title{Luminosity Functions of the Host Galaxies of Supernova}

\author[0009-0006-6622-5933]{Zhuoxi Liang}
\affiliation{Institute of Astronomy, Graduate School of Science, The University of Tokyo, 2-21-1 Osawa, Mitaka, Tokyo 181-0015, Japan}

\author[0000-0001-7266-930X]{Nao Suzuki}
\affiliation{E.O. Lawrence Berkeley National Laboratory, 1 Cyclotron Rd., Berkeley, CA, 94720, USA}
\affiliation{Department of Physics, Florida State University, 77 Chieftan Way, Tallahassee, FL 32306, USA}
\affiliation{Laboratoire de Physique Nucleaire et de Hautes-Energies, 4 Place Jussieu, 75005 Paris, France}

\author{Mamoru Doi}
\affiliation{Institute of Astronomy, Graduate School of Science, The University of Tokyo, 2-21-1 Osawa, Mitaka, Tokyo 181-0015, Japan}
\affiliation{Research Center for the Early Universe, Graduate School of Science, The University of Tokyo, 7-3-1 Hongo, Bunkyo-ku, Tokyo 113-0033, Japan}
\affiliation{Kavli Institute for the Physics and Mathematics of the Universe (WPI), The University of Tokyo Institutes for Advanced Study, The University of Tokyo, 5-1-5 Kashiwanoha, Kashiwa, Chiba 277-8583, Japan}

\author{Masayuki Tanaka}
\affiliation{National Astronomical Observatory of Japan, 2-21-1 Osawa, Mitaka, Tokyo 181-8588, Japan}
\affiliation{Department of Astronomical Science, The Graduate University for Advanced Studies, SOKENDAI, 2-21-1 Osawa, Mitaka, Tokyo, 181-8588, Japan}

\author{Naoki Yasuda}
\affiliation{Kavli Institute for the Physics and Mathematics of the Universe (WPI), The University of Tokyo Institutes for Advanced Study, The University of Tokyo, 5-1-5 Kashiwanoha, Kashiwa, Chiba 277-8583, Japan}

\begin{abstract}
We present the luminosity functions and stellar mass functions of supernova (SN) host galaxies and test if they differ from the functions of normal field galaxies. We utilize homogeneous samples consisting of 273 SNe Ia ($z\leq0.3$) and 44 core-collapse (CC) SNe ($z \leq 0.1$) from the Sloan Digital Sky Survey (SDSS) II Supernova Survey and the high-signal-to-noise-ratio photometry of galaxies from the Hyper Suprime-Cam Subaru Strategic Program (HSC SSP). SN hosts are classified into star-forming and passive galaxy groups based on the spectral energy distribution (SED) fitting. We find that the SN host luminosity functions and stellar mass functions deviate from those of normal field galaxies. Star-forming galaxies dominate the low-mass end of the SN Ia host mass function, while passive galaxies dominate the high-mass end. CC SNe are predominantly hosted by star-forming galaxies. In addition, intermediate-mass hosts produce CC SNe with the highest efficiency, while the efficiency of producing SNe Ia monotonically increases as the hosts become more massive. Furthermore, We derive the pseudo mass normalized SN rates (pSNuM) based on the mass functions. We find that the star-forming component of pSNuM$_{Ia}$ is less sensitive to the changes in stellar mass, in comparison with the total rate. The behavior of pSNuM$_{CC}$ suggests that the CC rate is proportional to the star-forming rate.

\end{abstract}


\defcitealias{2018PASP..130f4002S}{S18}
\section{Introduction}

Statistical properties of SN host galaxies serve as a proxy for understanding the correlation between the host environment of SN progenitors. It is widely considered that CC SNe, as the final fate of massive stars, are more associated with the star-forming activity compared to SNe Ia. Significant studies have been conducted on the correlation between SN rates (SNRs) and galactic properties. For example, SN Ia is the only SN that is found in all Hubble types, while CC SN is hosted by late-type galaxy \citep{2005PASP..117..773V}. Similarly, it has been revealed that CC SNR shows a stronger dependence of the Hubble types than that of SN Ia by a nearby complete sample of Lick Observatory Supernova Search \citep[LOSS, ][]{2011MNRAS.412.1473L}. 

On the other hand, recent cosmological studies utilizing SN Ia have proposed host stellar mass as a third parameter \citep{2010ApJ...715..743K}, in addition to the light curve shapes \citep{1993ApJ...413L.105P} and color \citep{1998A&A...331..815T}, for the standardization of SN Ia distance \citep{1995AJ....109....1H, 2010ApJ...722..566L, 2013ApJ...770..108C}, despite the question as to whether it is an intrinsic property or just an effect of dust \citep{2021ApJ...923..237J, 2022MNRAS.517.2360T}. Although it is not to be addressed here, exploring the star-formation-activity-related luminosity functions and mass functions may offer an alternative perspective.  \citet{2010AJ....139...39Y} produced luminosity functions of SN Ia hosts by utilizing 137 SNe Ia from SDSS II. However, the luminosity functions of CC SN hosts have not yet been discussed extensively.

This paper studies the luminosity functions of SN Ia and CC SN hosts in light of the star formation activities, using an unbiased SN catalog \citep[][hereafter S18]{2018PASP..130f4002S} from the untargeted SDSS-II Supernova Survey with a photometric limits of 0.3 for SNe Ia and 0.1 for CC SNe. Utilizing the galaxy photometry acquired from the Wide layer of Subaru Hyper Suprime-Cam Subaru Strategic Program  \citep[HSC-SSP,][]{2022PASJ...74..247A}, host galaxies of 30 SNe Ia that were classified as hostless in \citetalias{2018PASP..130f4002S} are newly identified. The SN host properties (stellar mass and star formation rate, SFR) are estimated by fitting their spectral energy distributions (SED).

This paper is organized as follows. In section \hyperref[sec:Data]{2}, we briefly introduce the samplings and photometry for SN host galaxies used in the analysis. Section \hyperref[sec:SED]{3} describes the SED fitting for SN hosts, through which we estimate the host properties. In Section \hyperref[sec:LF]{4}, the density distributions of SN host galaxy properties are derived. We discuss our results in Section \hyperref[sec:Discussion]{5}. The conclusion is given in Section \hyperref[sec:conclusion]{6}. Throughout this paper, we assume a $\Lambda$CDM model with cosmological parameters of $\Omega_{M}$=0.3, $\Omega_{\Lambda}$=0.7, and $H_{0}$=$\SI{70}{km\ s^{-1}\ Mpc^{-1}}$. All the magnitudes reported are in AB system.

\section{Data}
\label{sec:Data}
\subsection{SDSS-II Supernova Survey}
The three 3-month (2005-2007) SDSS-II Supernova Survey \citep{2008AJ....135..338F, 2008AJ....135..348S} was carried out on a 2.5 m telescope at Apache Point Observatory through \textit{ugriz} SDSS filters \citep{2010AJ....139.1628D}. The survey enables unbiased \citep{2008ApJ...682..262D} SN studies by repeatedly scanning the region, known as the Stripe 82, in the southern Galactic hemisphere from $-60^{\circ}<$ R.A. $< 60^{\circ}$, $-1.258^{\circ}<$ Decl. $< 1.258^{\circ}$ with an average cadence of four nights. Spectroscopic follow-ups were performed with a typical r-band limiting magnitude of $r \sim 21.5$ mag \citep{2012ApJ...755...61S}.

We begin with \citetalias{2018PASP..130f4002S} which provides by far the most homogeneous SN catalog. These SNe are classified into SN Ia, SN II, and SN Ibc based on the Bayesian probabilities, and more details can be found in \citetalias{2018PASP..130f4002S}. Given that the sample set of SNe Ibc is too small to conduct unbiased statistics, SNe II and SNe Ibc are designated as "CC SNe" hereafter. 

The host galaxy has also been identified for each SN when a minimum $d_{DLR}$ is obtained among host galaxy candidates, where $d_{DLR}$ is defined as the ratio of SN-galaxy separation to the directional light radius (DLR) \citep{2006ApJ...648..868S}. And $d_{DLR}$ is calculated in HSC r-band in this work. As it has been pointed out that the host galaxy selection maintains a high efficiency and less contamination from other sources when $d_{DLR} < 4$ in \citetalias{2018PASP..130f4002S}, we follow the same criterion in this work. 

In this SN catalog, 540 SNe Ia and 86 CC SNe are spectroscopically confirmed, 824 SNe Ia and 397 CC SNe are classified by photometric data with the spectroscopic redshift (\textit{spec-z}) of their host galaxies, and 624 SNe Ia and 1644 CC SNe are classified by photometric classification with the photometric redshift (\textit{photo-z}). As \textit{photo-z} brings undesirable uncertainty and degeneracy into SED fitting, we only adopt SNe with provided \textit{spec-z} in CMB frame, leaving 1364 SNe Ia and 483 CC SNe available.

\subsection{Subaru HSC SSP Data}
The HSC SSP survey is carried out on the 8.2 m Subaru Telescope in \textit{grizy} filters \citep{2018PASJ...70S...4A} with a limiting magnitude of 26.1 in r-band. The Wide layer scans over 1400 deg$^{2}$ though, therein only the region bounded by $-30^{\circ}<$ R.A. $ < 40^{\circ}$, $-1.25^{\circ}<$ Decl. $< 1.25^{\circ}$ is the overlap with the Stripe 82, leaving a final sky region of 175 deg$^{2}$ in this study. We employ the CModel photometry in the Wide layer \citep{2018PASJ...70S...6H}, which provides robust color information for extended sources.

Using high-signal-to-noise-ratio photometric data as input is key to perform accurate SED fitting, because a lower observational uncertainty enables a more valid estimation and a tighter constraint for the galaxy properties. For example, an accurate estimate of a galaxy's stellar mass can be acquired by combining the observational high-signal-to-noise-ratio luminosities and colors and the knowledge of a galaxy's mass-to-light ratio. Although photometry for SN host galaxies is available in SDSS, large uncertainties associated with high-redshift objects make it less desirable. Therefore, we instead make use of the data from the HSC SSP Wide layer which is capable of precise measurement within the redshift range of our interest.

We perform cross-matching and the galaxy data in HSC SSP Data Release 4 (DR4, in preparation) was retrieved within a separation of 1.0 arcsec (between the coordinates of identified galaxies reported by \citetalias{2018PASP..130f4002S} and those in HSC database). In order to provide a homogeneous input for the SED fitting, we also require all samples should have valid values in \texttt{cmodel\_[grizy]\_flux} and \texttt{cmodel\_[grizy]\_fluxerr} in the HSC Wide layer.  

\begin{deluxetable}{lllll}[htb!]
\tablecaption{Supernova Host Galaxies Identified in HSC-SSP Catalog}
\tablehead{
\colhead{SN class} & \colhead{S18}& \colhead{HSC}& \colhead{Selected}
}
\startdata
SNIa  & 222 & 27 & 133\\
SNIa? & 18 & 2 & 10\\
zSNIa  & 568 & 1 & 130\\
Ia total & ... & ... & 273\\
\hline\\
SNII  & 28 & 0 & 11\\
SNIb & 2 & 0 & 2\\
SNIc  & 4 & 0 & 2\\
zSNII  & 247 & 0 & 27\\
zSNIbc & 34 & 0 & 2\\
CC total & ... & ... & 44\\
\enddata
\tablecomments{Type category is described in \cite{2018PASP..130f4002S}. A prefix 'z' indicates that the host redshift is used as a prior in the classifications. Other classes are classified based on measured SN redshifts. 'SN Ia?' indicates that the a SN Ia is classified based on SN spectrum but the result is not conclusive. The columns 'S18' and 'HSC' show the number of SNe with host galaxies identified by \cite{2018PASP..130f4002S} and that of SNe with newly identified host galaxies by HSC (see Table \hyperref[tbl:hostless]{5}), respectively. The last column 'Selected' shows the number of SN host galaxies selected to derive luminosity function and mass function based on the redshift and luminosity selections.}

\end{deluxetable}

\label{tbl:SNsum}

\subsection{Photometry of low-z hosts}
In general, the dedicated software pipeline for HSC SSP is optimized for higher-redshift objects, but not for the low-redshift, very bright, and spatially over-extended ones -- the former ones are unlikely to have large spatial extensions in a bright level. For example, the HSC SSP uses small-size bins to estimate the sky background in the temporary over-subtraction \citep{2018PASJ...70S...5B}. In this study, it tends to underestimate the disk component of nearby galaxies that have large apparent sizes in the images, leading to biases in luminosity and colors at the low-redshift end. As redshift increases ($z > 0.1$), the measurements become more accurate. Based on the isophotal shape in photometry from the HSC catalog, we identified that the photometry of approximately $\sim30\%$ of the core-collapse hosts below redshift 0.1 was incorrect, which could introduce strong biases in inferring the galactic properties.

It motivates us to correct the photometric data of all nearby host galaxies of $z\leq0.1$. For each of the galaxies, we download the cutout image from HSC SSP S21A Wide database with a size of $52.92^2$ arcsec$^2$, centered at the SN location. We then use SExtractor \citep[][SEx hereafter]{1996A&AS..117..393B} to perform source detection and photometry for these host galaxies. For each image, we run SExtractor with a weighting map by variance, a seeing of FWHM=$0.7"$, a magnitude zero-point of 27 mag, and a background size of 21$^2$ arcsec$^2$. The effective gain for the image is 3.0 $\times$ N $e^-/$ADU, where N is the number of frames in the image stacking. 

The luminosity of each galaxy is then estimated by SEx's Model magnitude, which is comprised of a Sérsic spheroid (SPHEROID) profile and an exponential disk (DISK) profile. We would like to stress that the CModel magnitude in HSC is comprised of a de Vaucouleurs profile and a disk profile. Therefore, two different magnitude models compose the photometry of SN Ia hosts (SEx's model magnitude for $z\leq0.1$ hosts, and HSC's CModel magnitude for hosts at $0.1<z\leq0.3$). However, as the apparent size of the galaxies becomes smaller, background over-subtraction is negligible and these two magnitudes should be similar if they are in good agreement. We confirmed this consistency between these two magnitudes by using $z>0.3$ SN Ia host galaxies. In the case of CC SN hosts, only SEx's model magnitude is used because the redshift cutoff at $z\leq0.1$ is taken, as described in the next section. 

\begin{figure*}[htb!]
\label{fig:1}
\begin{center}
\includegraphics[width=18cm]{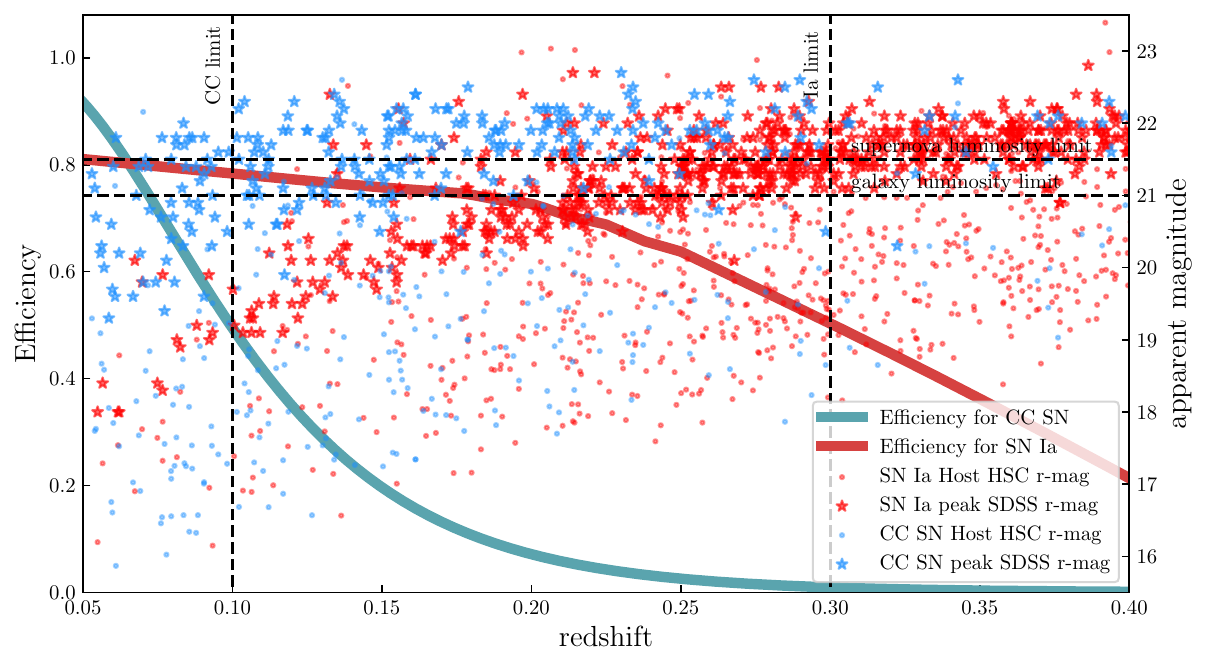} 
\end{center}
\caption{Efficiencies for SN Ia and CC SN as functions of redshift. The CC SN hosts within $z\leq0.1$ and the SN Ia hosts within $z\leq0.3$ are selected based on the estimated survey efficiency. To account for the incompleteness in sample brightness, the sample luminosity limit for SNe and that for their host galaxies are determined as $r\sim21.5$ mag and $r\sim21$ mag, respectively.}
\end{figure*}

\subsection{Efficiency}
\label{subsec:EFF}
In order to perform proper statistics, the weights of each SN and SN host should be considered. Traditionally, the survey efficiencies can be estimated by Monte-Carlo simulations \citep[e.g.][]{2008ApJ...682..262D} in the SNR studies. Nevertheless, the effective efficiencies can also be obtained backward, by making use of these reported SNRs
\begin{equation}
\begin{split}
    \varepsilon(z)&= \sum_{i}\frac{1+z}{r_{V}V_{C}T}H\left({\Delta z/2-|z_i-z|}\right)
\end{split}
\end{equation}
where $H$, $z_{i}$, $r_{V}(z)$, $T$, and $V_{C}(z)$ are the step function, the \textit{spec-z} of the i-th SN, the volumetric SNR, the total survey time for SNe in the observer frame (3 three-months), and the comoving volume of redshift bin $z\pm\Delta z/2$ corresponding to the sector that the HSC SSP footprints and the Stripe 82 overlap each other. We employ the reported SNRs (Ia: \citealt{2019MNRAS.486.2308F}; CC: \citealt{2014ApJ...792..135T}) to evaluate effective efficiencies for SNe Ia and CC SNe respectively. The efficiency curves are then interpolated by comparing to that in \citealt{2010ApJ...713.1026D} for SNe Ia and that in \citealt{2014ApJ...792..135T} for CC SNe, as shown as thick lines in Figure \hyperref[fig:1]{1}. 

The efficiency of SN Ia suggests that the sample set remains complete up to $z \sim 0.2$, which is in good agreement with previous estimates (\citetalias{2018PASP..130f4002S}; \citealt{2010AJ....139...39Y}). Since SDSS-II Supernova survey is designed and optimized for the SN Ia study in general, the efficiency for CC SN follows an unsurprising rapid decline as redshift increases. We adopt a redshift limit of 0.1 for CC SN, where efficiency decreases to half of the maximum value. We also set a lower redshift limit of $z_{\text{min}}=0.05$ in avoid of the additional uncertainties from local peculiar motions due to local clustering.

We also account for the completeness of SNe and SN hosts by studying their distribution along apparent magnitude in Figure \hyperref[fig:1]{1}. We first determine the limiting magnitude of SN detection of $r_{SN,lim}\sim21.5$ mag for this study, so that any SNe brighter than this limit will be included in the complete sample set. We note that this luminosity limit is the same as that in \citet{2010AJ....139...39Y}. We then consider the limit for SN Ia hosts by shifting the limiting magnitude (from $r\sim22.5$ mag to $20.5$ mag), until the statistics result becomes insensitive to the changing limit. It is found that the SN Ia hosts luminosity function remains almost unchanged from $r\sim21$ mag, suggesting that the SN Ia hosts samples are of high completeness. Thus, the SN Ia hosts limiting magnitude is determined as $r_{SN,lim}\sim21$ mag. Furthermore, the majority of CC SN hosts have apparent magnitudes that far exceed this limit due to their proximity. The same limiting magnitude is thus applied to CC SN hosts for consistency.

Table \hyperref[tbl:SNsum]{1} lists the number of classified SNe used in this work. The last column 'Statistics' refers to the SN hosts within selections by redshift and by SED fitting in \hyperref[sec:SED]{$\S$3}. They are used in \hyperref[sec:LF]{$\S$4}


\begin{deluxetable*}{lcl}[ht]
\tablecaption{CIGALE SED modeling parameters}
\tablehead{\colhead{Parameter}&&\colhead{Value}} 
\startdata
{     }&sfh2exp&{     }\\
\hline
$\tau_{main}$[Myr]&&10, 100, 500, 1000, 5000, 8000\\
$\tau_{burst}$[Myr]&&3, 5, 8, 15, 80\\
Age$_{main}$[Myr]&&3000, 5000, 11000\\
Age$_{burst}$[Myr]&&20, 200\\
$f_{burst}$&&0.0, 0.1, 0.2, 0.4\\
\hline
{     }&SSP \citep{2003MNRAS.344.1000B}&{     }\\
\hline
Initial Mass Function&&\citet{2003PASP..115..763C}\\
Metalicity&&0.02\\
\hline
{     }&Dust attenuation \citep{2000ApJ...533..682C}&{      }\\
\hline
E(B$-$V)$_{lines}$&{     }&0.005, 0.01, 0.025, 0.05, 0.1\\
&{     }&0.125, 0.15, 0.2, 0.4, 0.6\\
E(B$-$V)$_{factor}$&&0.25,0.5,0.75\\
\hline
{     }&Dust Emission \citep{2014ApJ...784...83D}&{     }\\
\hline
$\alpha$&&2.0\\
\enddata
\end{deluxetable*}

\label{tbl:param}

\section{SED Fitting}
\label{sec:SED}
In order to infer the galaxy properties for SN hosts, we employed CIGALE \citep{2019A&A...622A.103B} to perform SED fitting with the HSC photometry as well as the host \textit{spec-z} for all the 808 SNe Ia hosts and 315 CC SNe hosts. CIGLAE provides us with a self-consistent framework where the galaxy SED is obtained, and the SED model parameters, including stellar mass $M_{\star}$ and SFR, are estimated simultaneously for each SN host. SN host galaxies' properties are summarized in Table \hyperref[tbl:main]{3}.

\subsection{Choice of Priors}
We here briefly review the modeling modules. We assume double-exponential star formation histories (\texttt{sfh2exp}) and spectra template from \texttt{BC03} \citep{2003MNRAS.344.1000B} with the \citet{2003PASP..115..763C} initial mass function (IMF), which have been widely and intensively used in previous studies. Fitting parameters are shown in Table \hyperref[tbl:param]{2}.

\begin{figure*}[ht!]
\label{fig:SEDex}
\begin{center}
\includegraphics[width=18cm]{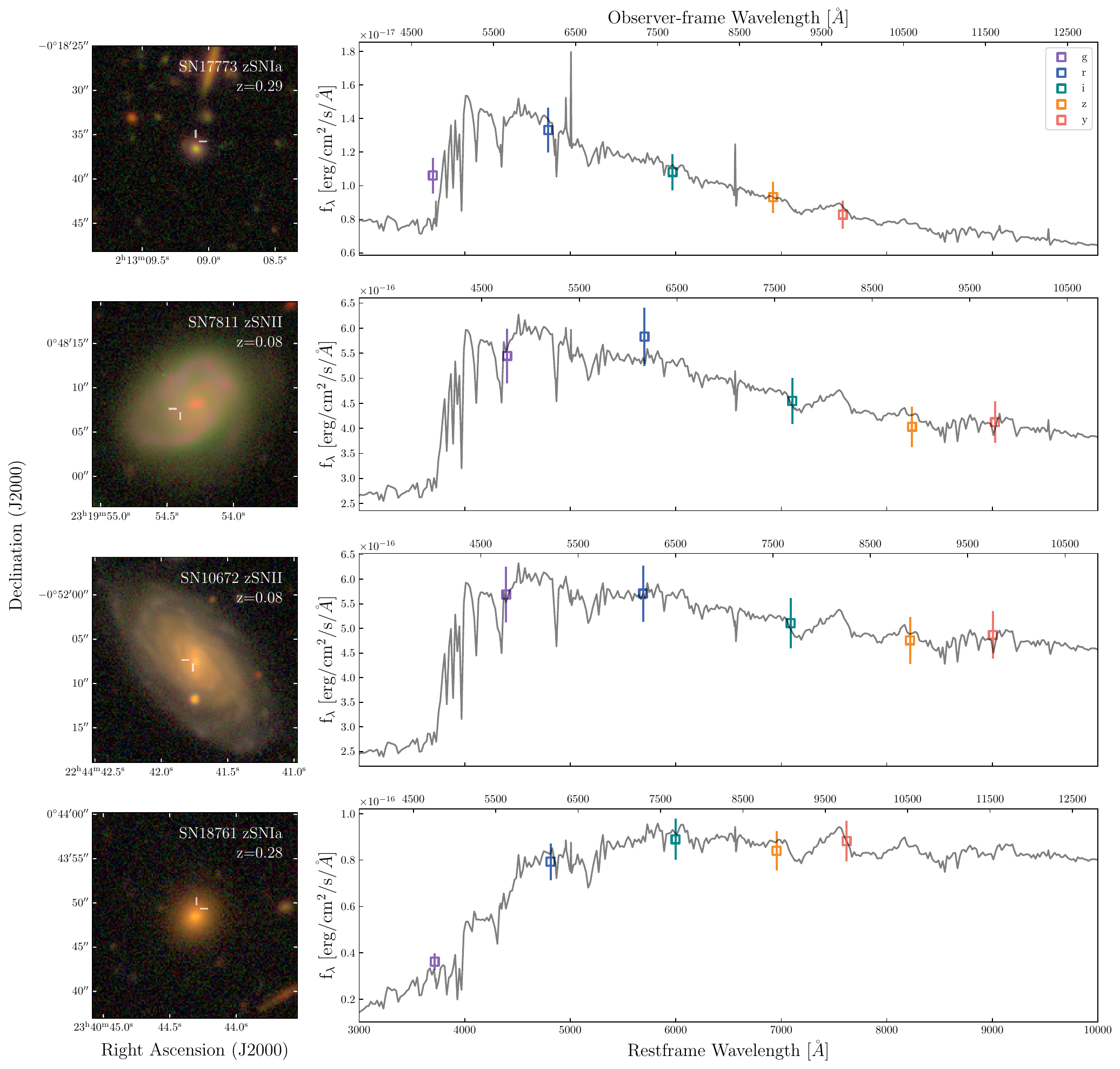} 
\end{center}
\caption{Exemplary SED fittings for SN host galaxies. \textit{left}: the archival HSC cutout images of SN hosts with color mixtures of HSC WIDE g, r, and i-band through blue, green, and red channels. \textit{right}: restframe SED models of SN hosts obtained from CIGALE. Colored squares mark the HSC observations with 1-$\sigma$ errors in each HSC filter band.}
\end{figure*}

In this study, we exclude high attenuation models where E(B-V)$_{lines}>0.6$. The lack of UV and IR photometry makes it difficult to predict accurate dust attenuation for SN hosts. Modeling with high attenuation parameters will lead to overestimations of SFR for most of the SN hosts, resulting in a systematical bias. However, it should be kept in mind that rare extreme cases do exist. Since dust emission is poorly constrained without infrared data, we simply set the radiation field intensity power-law index $\alpha=2$. Likewise, the radio module and the AGN module are not included throughout our modeling due to the lack of detection, and the AGN activities are not significant in the low-z universe. Figure \hyperref[fig:SEDex]{2} displays the SED results for 4 representative SN host galaxies. It shows that SED models are consistent with the galaxy morphological types as well as colors.

\subsection{Starforming Galaxies vs Passive Galaxies}
\label{sec:SFB}
\begin{figure*}[ht]
\label{fig:SFB}
\begin{center}
\includegraphics[width=18cm]{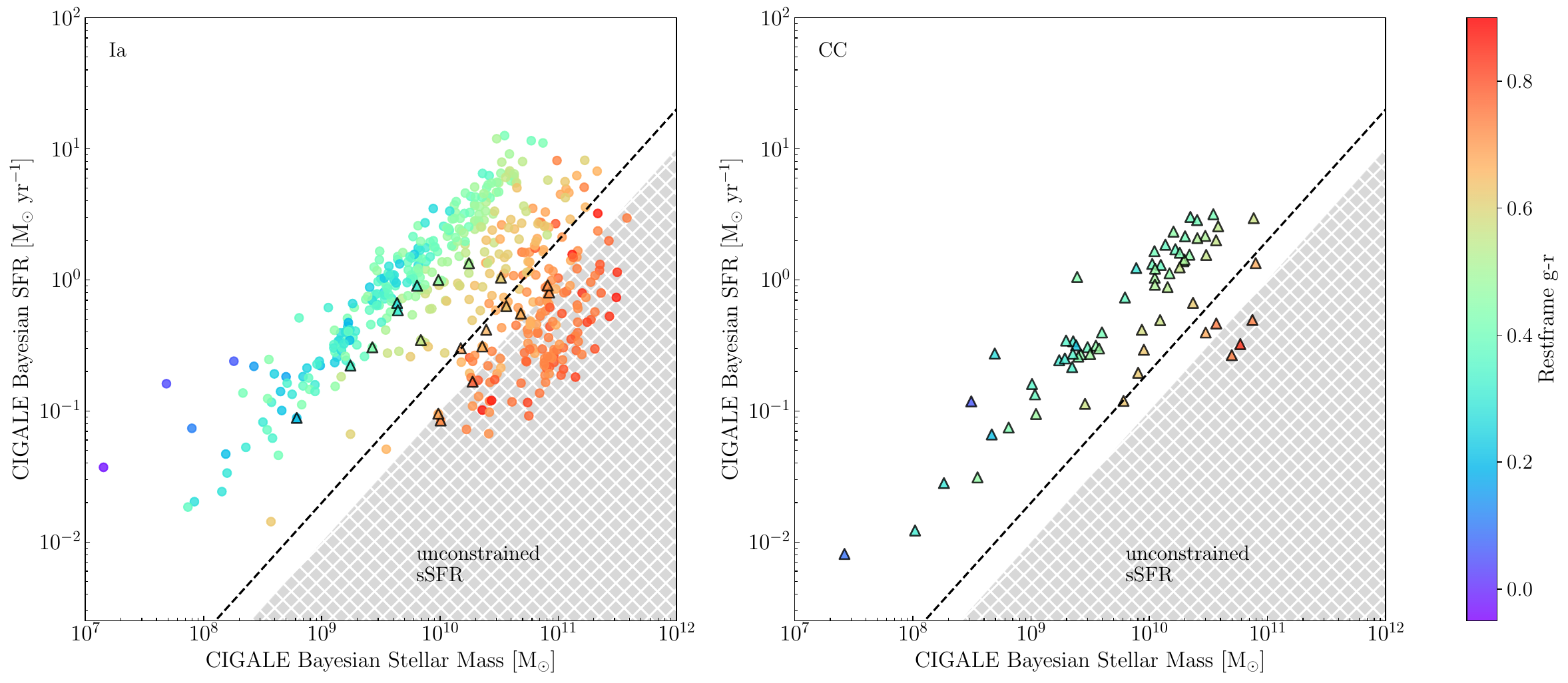} 
\end{center}
\caption{The results SFR versus the stellar mass diagram for SN Ia and CC SN host galaxies within redshift selections, while luminosity cut is not applied. Each scatter marks one galaxy in our sample set without volumetric density information. Colored dots represent the SN hosts at $0.1\leq z<0.3$, and colored triangles represent the SN hosts at $z < 0.1$. Each marker is colored by the galaxy restframe g-r reported by CIGALE. The dashed line corresponds to the boundary between star-forming galaxies and passive galaxies with a log(sSFR/yr$^{-1}$) $= -10.6$. The estimated SFR becomes unconstrained when log(sSFR/yr$^-1$)$\lesssim -11$ by optical SED fittings \citep{2013ApJ...770..107C}.}
\end{figure*}

Figure \hyperref[fig:SFB]{3} shows SFR - stellar mass diagram for the SN hosts using the SED fitting results. For each galaxy, galaxy properties are estimated by weighting all the possible values for different SED models by the Bayesian-like likelihood. Readers are referred to \citet{2019A&A...622A.103B} for detailed descriptions. In addition, we exclude bad estimations where even the best SED models have a reduced $\chi^2$ (i.e., $\chi^2/d.o.f$) larger than 5.

Figure \hyperref[fig:SFB]{3} shows a bimodal distribution for SN Ia hosts: the concentration on star-forming main sequence, and a cluster at the passive region around log$(M/M_{\odot})\sim11$. This pattern is described in \citet{2015ApJ...801L..29R} but for nearby galaxies. Meanwhile, most of the CC SN hosts are located at the star-forming main sequence, and it shows no evidence of a cluster pattern in the passive region. This suggests that CC SN events are preferentially hosted by young galaxies, which is in agreement with previous studies \citep{2005PASP..117..773V, 2022ApJ...927...10I}. 

Based on the distribution pattern of the whole sample set of SN Ia hosts and CC hosts, we also define star-forming (passive) galaxies as those with sSFR higher (lower) than $10^{-10.6}$ yr$^{-1}$. Seven passive hosts are (5 were used for statistics after luminosity selection) for CC SN, but they are likely star-forming galaxies within SED fitting uncertainties, see Appendix \hyperref[sec: A2]{2} for more discussions. We also note that our classification is simply based on the SED fitting estimation, and thus potentially biased by the observational inputs. As we only include optical photometries in the fitting, the estimated sSFR in general correlated to restframe colors.


\begin{deluxetable*}{cccccccccc}[ht]
\tablecaption{SDSS supernova host galaxies catalog}
\tablehead{
\colhead{SNID}&
\colhead{z$_{\text{CMB}}$}&
\colhead{Class}&
\colhead{$\alpha^{\text{Host}}_{J2000}$}&
\colhead{$\delta^{\text{Host}}_{J2000}$}&
\colhead{m$_r$}&
\colhead{d$_{\text{DLR}}$}&
\colhead{log$M/M_\odot$}&
\colhead{M$_g$$-$M$_r$}&
\colhead{M$_r$}\\
\colhead{ }&
\colhead{ }&
\colhead{ }&
\colhead{(deg)}&
\colhead{(deg)}&
\colhead{(mag)}&
\colhead{ }&
\colhead{ }&
\colhead{(mag)}&
\colhead{(mag)}
}
\startdata
703 & 0.296 & zSNIa & 336.2180 & 0.6508 & 20.80 & 0.81 & 10.06 & 0.49 & -20.34 \\ 
739 & 0.106 & SNIa & 14.5958 & 0.6794 & 16.89 & 3.22 & 11.03 & 0.90 & -21.73\\ 
762 & 0.190 & SNIa & 15.5361 & -0.8797 & 17.43 & 4.54 & 11.10 & 0.70 & -22.61 \\ 
770 & 0.041 & zSNII & 31.6602 & 0.6578 & 17.21 & 0.26 & 9.55 & 0.47 & -19.12  \\ 
774 & 0.092 & SNIa & 25.4637 & -0.8767 & 17.09 & 1.72 & 10.68 & 0.89 & -21.16 \\ 
779 & 0.237 & zSNIa & 26.6738 & -1.0206 & 19.56 & 0.10 & 10.12 & 0.58 & -21.00 \\ 
859 & 0.277 & zSNIa & 350.5517 & 0.3866 & 21.49 & 0.63 & 9.45 & 0.39 & -19.35 \\ 
863 & 0.093 & zSNII & 354.6942 & 0.2842 & 17.96 & 0.23 & 10.05 & 0.56 & -20.27 \\ 
911 & 0.206 & zSNIa & 38.6910 & -0.1153 & 19.65 & 0.94 & 10.08 & 0.46 & -20.51 \\ 
986 & 0.279 & zSNIa & 352.8123 & 1.1416 & 19.99 & 0.72 & 10.13 & 0.52 & -21.02 \\ 
... & ... & ... & ... & ... & ... & ... & ... & ... & ... \\
\enddata
\end{deluxetable*}

\label{tbl:main}

\section{Luminosity FunctionS AND MASS FUNCTIONS}
\label{sec:LF}
To correct the Malmquist bias \citep{1922MeLuF.100....1M}, we employ $1/V_{\text{max}}$ method \citep{1968ApJ...151..393S} to perform statistics for our SN hosts luminosity. Explicitly, the maximum comving volume to find the \text{i-th} SN 

\begin{equation}
    V(z)_{\text{max, i}} = \frac{\Omega}{4\pi}\int_{0.05}^{z_{\text{max, i}}}\frac{dV}{dz}dz
\end{equation}
where $\Omega$ is the overlap area between the HSC SSP footprints and the Stripe 82. The maximum volume of the detection, encompassing both the SN and galaxy, is determined by $z_{\text{max, i}}$, which is the minimum value between $z_{\text{host,max,i}}$ and $z_{\text{SN,max,i}}$, for the $i$-th SN host. Here, $z_{\text{host,max,i}}$ is determined by the model SED. In order to determined the $z_{\text{SN,max,i}}$ for each SN Ia, we calculate the apparent r-band magnitude (at the brightest epoch) of each redshift by \texttt{SNCosmo} \citep{2016ascl.soft11017B} using reported SALT2 parameters in \citetalias{2018PASP..130f4002S}. When there is no valid SALT2 parameters, $z_{\text{SN,max,i}} = 0.3$ is assumed. On the other hand, the K-corrections for CC SNe are negligible due to their proximity and their lumionsities being close to the SN luminosity limit, as shown in Figure\hyperref[fig:1]{1}. Therefore, the $z_{\text{SN,max,i}}$ is simply determined by distance modulus in the case of CC SN. In these procedures, the luminosity limits are the sample limits described in Section \hyperref[subsec:EFF]{2.4}. 

The luminosity density $\phi(M_{^{0.1}r})$, where $M_{^{0.1}r}$ is the r-band absolute magnitude $M^{0.1r}$ at z=0.1 frame, can then be evaluated through summing over and giving weights to the number of SN hosts which fall within the mag bin $M_{^{0.1}r}\pm\Delta M_{^{0.1}r}/2$:

\begin{equation}
\label{eq:3}
\begin{split}
    \phi(M_{^{0.1}r})\Delta M_{^{0.1}r} = 
    \frac{1+z}{T}\sum_{\substack{|M_{^{0.1}r}-M_{^{0.1}r,i}|\\\leq\Delta M_{^{0.1}r}/2}}\frac{1}{\varepsilon(z_i)V(z_i)_{\text{max}}}
\end{split}
\end{equation}
where T is the survey time (270 days for SDSS-II SN survey), $z$ is the mean redshift of the SN samples ($z=0.22$ for SN Ia and $z=0.08$ for CC SNe), and $\varepsilon(z_i)$ is the survey efficiency at SN redshift $z_{i}$ estimated in Figure \hyperref[fig:1]{1}. We apply redshift selection $z < 0.3$ for SN Ia, where efficiency declines to half of that at $z = 0.05$. Similarly, our evaluation suggests that the sample set of CC SNe still enable us to conduct unbiased statistics below $z = 0.1$. In total, 273 SNe Ia and 44 CC SNe, and their host galaxies are employed for statistics. Likewise, we calculate the stellar mass function by equation \hyperref[eq:4]{4}. Additionally, galaxies are further separated into star-forming and passive subsets. Finally, the upper and lower limits are evaluated through Poisson statistics \citep{1986ApJ...303..336G}.

The number density distribution for luminosity and $M_{\star}$ can be empirically described by the Schechter function \citep{1976ApJ...203..297S}. In practice, it is more convenient to evaluate the number density in terms of absolute magnitude:

\begin{equation}
\label{eq:4}
\phi\ dM_{^{0.1}r} = 0.4\ln{10}\ \phi^{*}\exp{[-10^{0.4(M_{^{0.1}r}-M^{*})}]}\left[10^{0.4(M_{^{0.1}r}-M^{*})}\right]^{\alpha+1} dM_{^{0.1}r}
\end{equation}
which is determined by a faint-end slope $\alpha$, a characteristic absolute magnitude $M^{*}$, and a normalization factor $\phi^{*}$ for the number density at $M^{*}$. Likewise, the calculation of mass function $M_{\star}$ is described by:
\citep{2016MNRAS.459.2150W}:

\begin{equation}
\label{eq:5}
\phi_{\star}\ d\log{M_{\star}} = \ln{10}\ \sum_{i=1,2} \exp{[-10^{\log{M_{\star}}-\log{M_{\star, i}^{*}}}]}\left(\phi_{\star, i}^{*}10^{\log{M_{\star}}-\log{M_{\star}^{*}}}\right)^{\alpha_i+1} d\log{M_{\star}}
\end{equation}
where parameters are analogous to those of $M^{0.1r}$ but now for the double-component stellar mass function. Since the number of SN hosts is limited in this work, we adopt a single-component fitting in avoid of over-fitting, despite the fact that recent studies have found double Schechter function to be a better description \citep[][and references therein]{2016MNRAS.459.2150W}. 

\begin{deluxetable*}{lccc}[ht]
\tabletypesize{\scriptsize}

\tablecaption{Schechter function parameters}
\tablehead{\colhead{Parameters}&\colhead{Field Galaxy$^{a,b}$}&\colhead{SN Ia Host}&\colhead{CC SN Host}} 

\startdata
{     }&Luminosity Function&{     }\\
\hline
$\phi^{*}$ [$h^{3}$Mpc$^{-3}$mag$^{-1}$yr$^{-1}$] & $1.49\times10^{-2}$ & $(5.37\pm1.80)\times10^{-5}$ & $(3.14\pm0.94)\times10^{-4}$\\
$M^{*}-5\log{h}$ [mag]&$-20.44$ & $-20.60\pm0.28$ & $-19.01\pm0.25$\\
$\alpha$&$-1.05$&$-0.62\pm0.18$&$-0.19\pm0.30$\\
\hline
{     }&Mass Function&{     }\\
\hline
$\phi^{*}_{\star,1}$ [$h^{3}$Mpc$^{-3}$dex$^{-1}$yr$^{-1}$] & $9.78\times10^{-3}$ & $(4.80\pm0.97)\times10^{-5}$&$(4.26\pm2.17)\times10^{-4}$\\
$\phi^{*}_{\star,2}$ [$h^{3}$Mpc$^{-3}$dex$^{-1}$yr$^{-1}$] & $4.90\times10^{-4}$ &--&--\\
log($M^{*}_{\star}$/$M_{\odot}$) [dex] & $10.79$ & $11.00\pm0.05$ & $10.36\pm0.14$\\
$\alpha_{1}$&$-0.79$&$-0.606\pm0.08$&$-0.067\pm0.30$\\
$\alpha_{2}$&$-1.69$&--&--\\
\enddata
\tablenotetext{a}{h = 1 in \cite{2003ApJ...592..819B} and 0.7 in \cite{2016MNRAS.459.2150W}.}
\tablenotetext{b}{The normalization parameters $\phi^{*}$ and $\phi_{\star}^{*}$ have units of [$h^{3}$Mpc$^{-3}$mag$^{-1}$] and [$h^{3}$Mpc$^{-3}$dex$^{-1}$] in normal field galaxy's functions.}
\end{deluxetable*}

\label{tbl:LF_param}

\begin{figure*}[htb!!]
\centering

\gridline{
        \fig{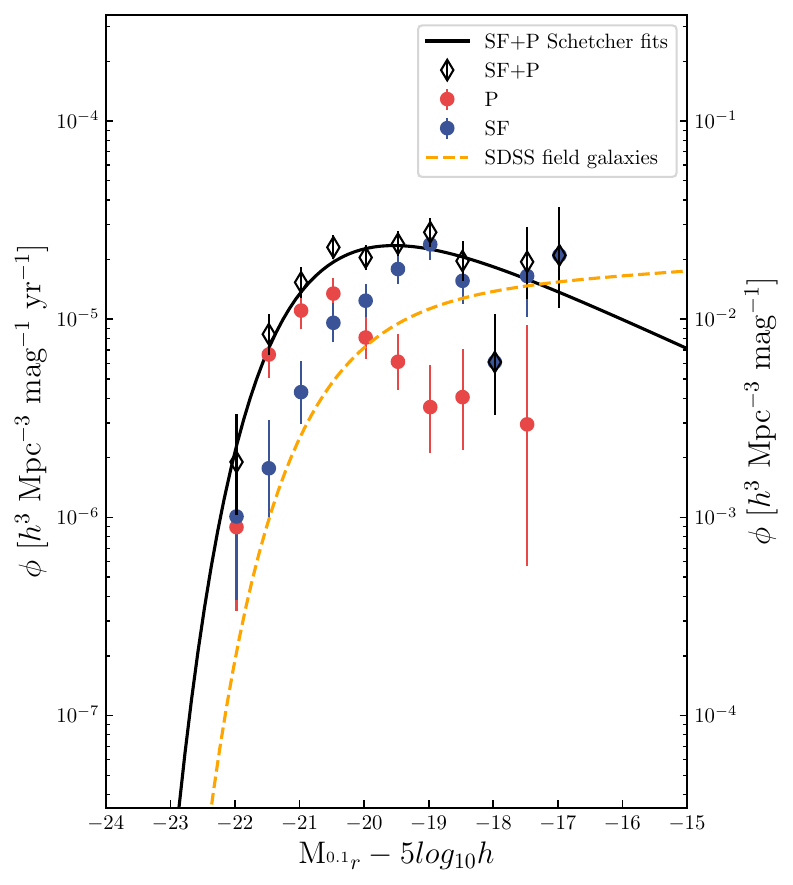}{0.5\textwidth}{(a)\label{fig:4a}}
        \fig{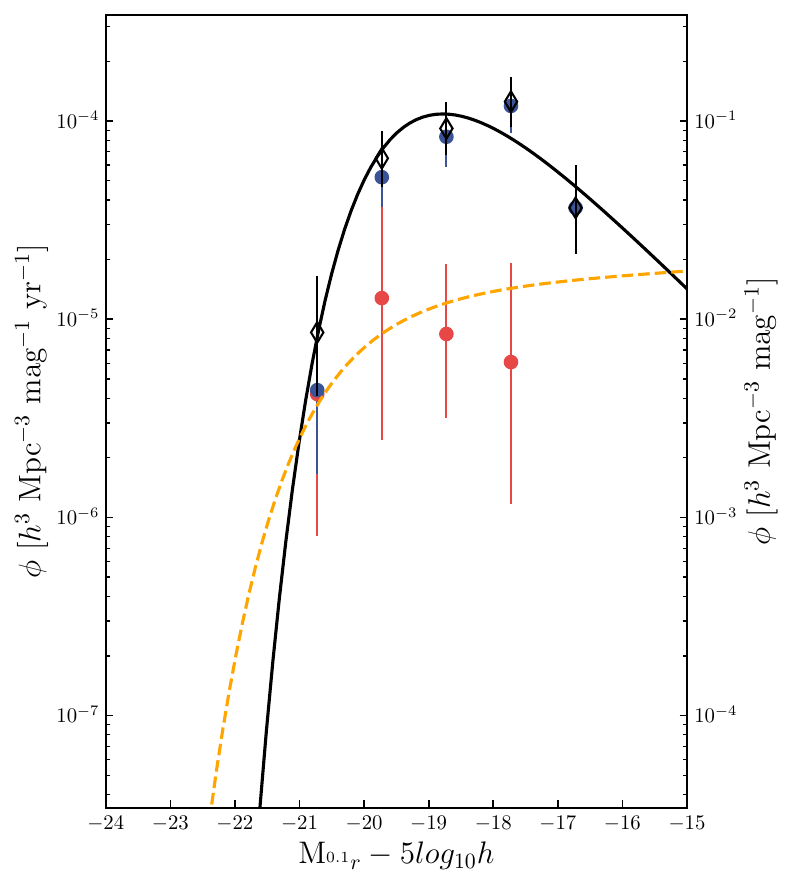}{0.5\textwidth}{(b)\label{fig:4b}}}

\caption{The luminosity function of Ia (a) and CC (b) host galaxies is represented by black, blue, and red circles, indicating the density of all galaxies, star-forming galaxies, and passive galaxies, respectively. The Schechter fitting of all galaxies is depicted by the black curve. And the SDSS field galaxy is marked in the orange dashed curve \citep{2003ApJ...592..819B} with the right axis.}
\label{fig:4}
\end{figure*}

\begin{figure*}[htb!]
\label{fig:5}
\centering
\gridline{
        \fig{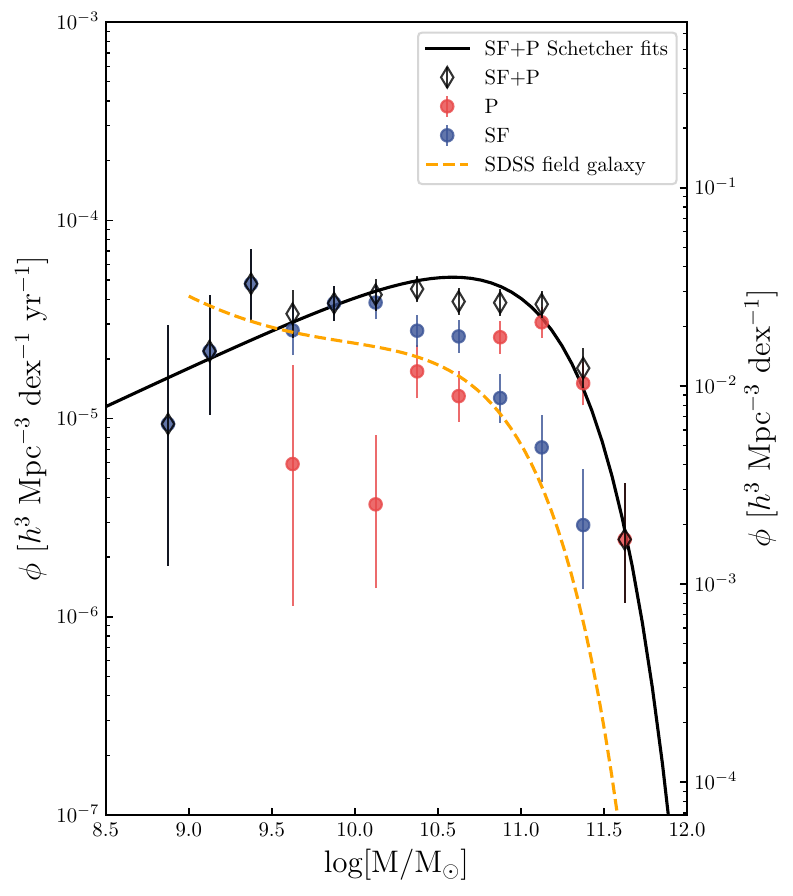}{0.5\textwidth}{(a)\label{fig:5a}}
        \fig{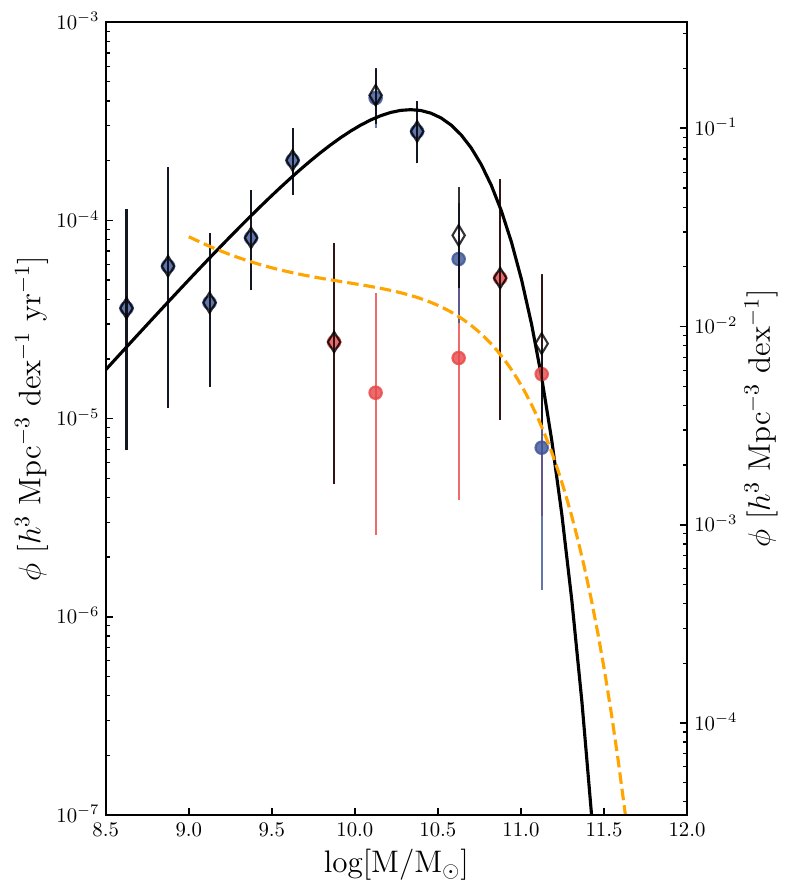}{0.5\textwidth}{(b)\label{fig:5b}}}

\caption{Same as Figure \hyperref[fig:4]{4} but for stellar mass functions of Ia (a) and CC (b) host galaxies. The orange dashed curve denotes the mass function of field galaxy reported in \cite{2016MNRAS.459.2150W}.}
\end{figure*}

\subsection{Results}
Figure \hyperref[fig:4]{4} and Figure \hyperref[fig:5]{5} display the derived luminosity functions and stellar mass functions for SN Ia and CC SN hosts, respectively. The fitting parameters of these functions are summarized in Table \hyperref[tbl:LF_param]{4}.

The fitting parameters show that the CC SN host is $\sim 1.6$ mag fainter than SN Ia host, and is $\sim 1.4$ mag fainter than SDSS Field galaxy in terms of characteristic $M^{*}$. 
Functions of CC SN hosts present deficiencies at their bright ends by a significant number of galaxies, compared to the functions of general field galaxies. Since galaxy stellar mass is proportional to luminosity, we found consistent conclusions for stellar mass functions.

Figure \hyperref[fig:4]{4} and Figure \hyperref[fig:5]{5} also show that SDSS field galaxy (of comparable redshift ranges) has a flat-distribution in the faint tail in the luminosity function, and an up-tilted faint tail in the stellar mass function. However, within 1-$\sigma$ uncertainty, no evidence of such faint-end distribution is found for SN Ia and CC SN hosts in their luminosity functions and stellar mass functions. 

By comparing the normalization factor $\phi^{*}$, the number density per year of SN Ia hosts are about 1 out of 277 field galaxies and that of CC SN hosts are 1 out of 47 field galaxies (at each characteristic luminosity).

The luminosity functions and stellar mass functions are further divided into star-forming and passive group according to Section \hyperref[sec:SFB]{3.2}. It is shown that passive SN Ia hosts dominate the bright (massive) end, while star-forming hosts dominate the faint (low-mass) end, following a general rule of galaxies. The volumetric densities of these two groups are comparable at $M_{r}^{0.1} \sim -20$ mag, or at log$(M/M_{\odot}) \sim 10.5$. The passive group of SN Ia hosts dominates the aforemetioned excess SN Ia rate per galaxy at the massive/bright end. On the other hand, CC SN hosts are dominated by star-forming galaxies at all magnitude. Unlike the SN Ia case, no evidence shows passive galaxies dominate the massive (bright) end.

\section{Discussion}
\label{sec:Discussion}

\subsection{Supernova Yield Efficiencies}
\label{sec:SNYE}

It is clear that being weighted by the SN volume and the galaxy volume, the host galaxy's stellar mass function and luminosity function deviate from that of field galaxy's. The difference between the functions of field galaxy and those of SN hosts can mainly be attributed to the SNR per galaxy. Both luminosity function and stellar mass function indicate that a more massive, brighter galaxy produces more SNe Ia, suggesting that SN Ia has a stronger dependence on stellar mass than on SFR. While intermediate-mass galaxies have the highest rates of CC SNe events, likely a result of the balance between SFR and stellar mass. Our result is in agreement with previous findings \citep{2010AJ....139...39Y,2011MNRAS.412.1473L}.

\begin{figure}[htb!]
    \includegraphics[width = 0.98\linewidth]{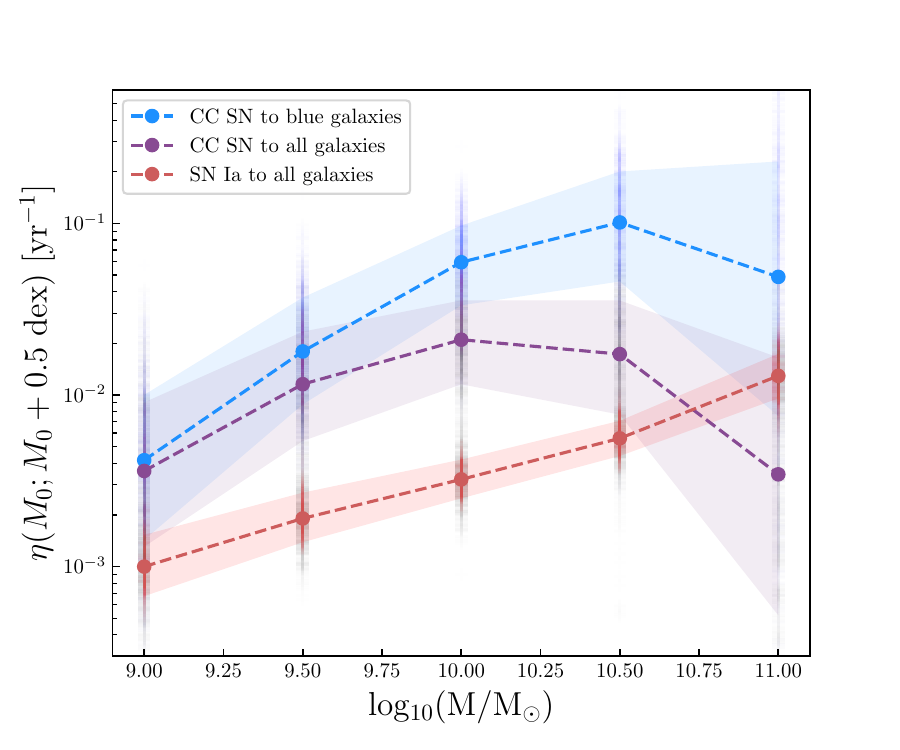}
    \caption{SN Ia and CC SN yield efficiency as a function of the lower bound of integral $M_{0}$ with an interval length of 0.5 dex. SN Ia is compared to all the field galaxies (red), while CC SN is compared to all (purple) and blue population (blue) galaxies, respectively. Each scatter mark an MC experiment result. Filled circles show the MC median values, lines represent the linear interpolation, and the shaded regions show the 16-84 percentile of the MC results.}
    \label{fig:6}
\end{figure}

\begin{figure}[htb!]
    \includegraphics[width = 0.98\linewidth]{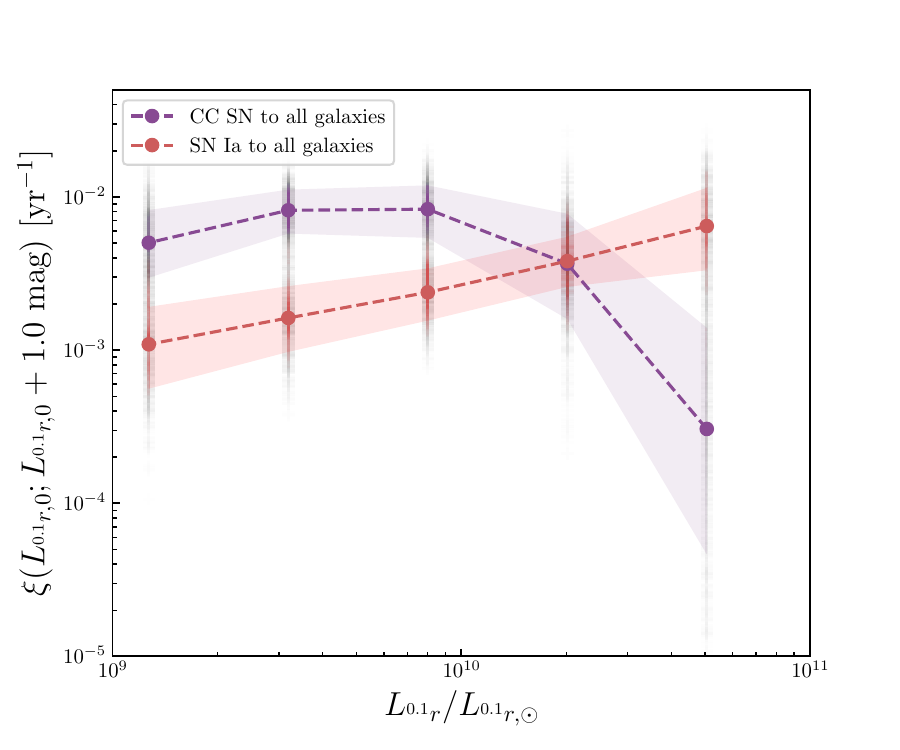}
    \caption{Same as Figure \ref{fig:6} but weighted by luminosity with an interval length of 1.0 mag. Luminosity is normalized by the r-band absolute solar magnitude $M_{^{0.1}r,\odot}=4.76$ \citep{2003ApJ...592..819B}.}
    \label{fig:7}
\end{figure}

To further explore the SN preferences on field galaxies, we define the dimensionless SN yield efficiency by comparing the integral of mass of SN hosts and that of field galaxies:

\begin{equation}
    \eta(M_0, M_1) = \frac{\int_{M_0}^{M_1}M\phi^{SN}(M)dM}{\int_{M_0}^{M_1}M\phi^{gal}(M)dM}
\end{equation}
(Mass-weighted)

\begin{equation}
    \xi(L_0, L_1) = \frac{\int_{L_0}^{L_1}L\phi^{SN}(L)dL}{\int_{L_0}^{L_1}L\phi^{gal}(L)dL}
\end{equation}
(Luminosity-weighted)

where $\phi^{SN}$ is the SN host mass function, and $\phi^{gal}$ is the field galaxy mass function of a comparable redshift range (\citealt{2016MNRAS.459.2150W} here). To account for the effects of the uncertainty, we estimate $\eta$ through Monte Carlo (MC) approaches given the results in Table \hyperref[tbl:LF_param]{4}. For both CC SN and SN Ia, $1,000$ randomised SN and field galaxy luminosity functions are generated, where Gaussian distributions are assumed for all parameters except $\phi_0$. Gamma distributions are assumed for $\phi_0$ as normalization factors should be positive values. Figure \hyperref[fig:6]{6} and Figure \hyperref[fig:7]{7} show our MC result $\eta$ as a function of $M_0$ with an integral interval width of 0.5 dex and $\xi$ as a function of $L_0$ with a width of 1.0 mag.


\subsubsection{SN Ia Efficiency}
\label{sec:SNYE_Ia}

From the yield efficiencies, we find that $\eta_{\text{Ia}}$ and $\xi_{\text{Ia}}$monotonically increase as the host galaxies get more massive and more luminous, indicating that stellar mass plays a dominating role in producing SNe Ia. Their tendencies are in good agreement with \cite{2021MNRAS.506.3330W} which reports a monotonically increasing SNR (per galaxy) toward the massive end, despite the different redshift selections. Meanwhile, the role of instant SFR remains elusive, probably because the wide range of SN Ia progenitor delayed time \citep{2010ApJ...722.1879M,2011MNRAS.417..408R} weakens their relations. 

The monotonically increasing trend of SN Ia efficiency (the red curve in Figure 6) can be attributed to the fact that SN Ia events can happen in passive galaxies. In general, SNR$_{\text{Ia}}$ is parametrized by a two-component model \citep[e.g., ][]{2005ApJ...629L..85S} which contains a prompt population (from the young stellar population) and a delayed population (from the old stellar population) of SNe Ia.  Since passive galaxies dominate the massive end of the SN Ia host mass function, the overall increasing trend suggests that the rising mass-integrated rate of delayed population SN Ia in passive hosts should have offset the declining rate attributed to the depletion of star-forming hosts in the massive end.

\subsubsection{CC SN Efficiency}
\label{sec:SNYE_CC}

On the other hand, $\eta_{\text{CC; all}}$ peaks at intermediate massive host ($\sim 10^{10}M_{\odot}$), and $\xi_{\text{CC; all}}$ peaks at intermediate luminous host ($\sim 10^{10}L_{^{0.1}r,\odot}$). The deficiency in the massive end is expected, because the presence of a sequence of passive galaxies beyond $10^{10} M_{\odot}$ \citep{2019MNRAS.482.1557S} significantly reduces $\eta_{\text{CC; all}}$. As star-forming galaxies predominantly hosts CC SNe, we further exclude the red population field galaxies \citep{2012MNRAS.421..621B}, and the resulted $\eta_{\text{CC; blue}}$ shows a stronger dependence on stellar mass. However, the slope of $\eta_{\text{CC}}$ decreases towards the massive end in either case, indicating a complex combination of impacts from galaxy properties. Possible causes include but not limited to that CC SNe follow star-forming activities immediately so that current properties, such as the star-forming component of stellar mass, of galaxy play more important roles. Moreover, the yield efficiencies serve as good references for observers to find SNe by monitoring specific kinds of galaxies. It clearly shows that it achieves the highest efficiency to find CC SNe by monitoring intermediate massive galaxies, and to find SNe Ia by monitoring the most massive galaxies.

The decreasing CC SN efficiency (the purple curve) at the massive end is due to that CC SNe hardly occur in passive galaxies. In addition, the CC SN efficiency becomes steady at the massive end when the it is only weighted by the blue population galaxies (the blue curve). If it is assumed that the CC SN rate is proportional to the SFR, this trend is consistent with the reported SFR-stellar mass relation in the local universe, which shows a turning beyond $10^{10.5}\ M_{\odot}$ \citep[][ and the references therein]{2020MNRAS.496.3668D}.

\subsection{Pseudo Supernova Rate per Stellar Mass}
\label{sec:pSNuM}
Besides the SN efficiencies, our estimated stellar mass functions also contain the SNR information which provide a proxy to estimate the pseudo SNR per stellar mass (pSNuM):

\begin{equation}
    \text{pSNuM}(M_0, M_1) = \frac{\int_{M_0}^{M_1}\phi^{SN}(M)dM}{\int_{M_0}^{M_1}M\phi^{gal}(M)dM}
\end{equation}

The resulting pSNuMs for SN Ia and CC SN are shown in Figure \hyperref[fig:pSNuM]{8} with an integral width of 0.25 dex in stellar mass. 

\subsubsection{SN Ia pSNuM}
\label{sec:pSNuM_Ia}

The $\text{pSNuM}_{\text{Ia}}$ follows a decreasing power-law below $10^{10.5} M_{\odot}$ and then become flat toward the massive end. Moreover, the $\text{pSNuM}_{\text{Ia}}$ is consistent with the predictions from the delayed-time distribution model from Dark Energy Survey \citep[DES][]{2021MNRAS.506.3330W}.  

The power-law behavior of $\text{pSNuM}_{\text{Ia}}$ is consistent with the literature \citep{2011MNRAS.412.1473L,2013MNRAS.430.1746G}.  To further explore the stellar mass dependence, we decoupled the $\text{pSNuM}_{\text{Ia}}$ into star-forming group $\text{pSNuM}_{\text{Ia, SF}}$ and passive group $\text{pSNuM}_{\text{Ia, P}}$by assuming that the faint-end slope of star-forming component (see Appendix for the details).  The estimated results are shown in Figure \hyperref[fig:pSNuMIa]{9}. The $\text{pSNuM}_{\text{Ia}}$ has a flat distribution in star-forming hosts and a complicated evolution in passive hosts, and overall the $\text{pSNuM}_{\text{Ia}}$ is higher in star-forming hosts than passive hosts. 

As star formation are overall similar in producing stars, the mild evolution of $\text{pSNuM}_{\text{Ia}}$ in star-forming galaxies qualitatively suggests that the star-forming stellar mass plays a leading role in determining the behavior of $\text{pSNuM}_{\text{Ia, SF}}$. It is found that $\text{pSNuM}_{\text{Ia, SF}}$ declines slower towards the massive end in comparison with that of $\text{pSNuM}_{\text{CC}}$. Since $\text{pSNuM}_{\text{CC}}$ is a potential tracer of star-formation rate per stellar mass (see Section \hyperref[sec:pSNuM_CC]{5.2.2}), it is likely that the slow-declining tail of $\text{pSNuM}_{\text{Ia, SF}}$ is the result of that SNe Ia are produced by both young and ole population stars, with overall longer delayed times. The mild evolution may also come from other mass-depending factors including the general metallicity-mass relation. 

On the other hand, $\text{pSNuM}_{\text{Ia, P}}$ presents a complicated evolution along stellar mass, which is likely a result of the long-delayed time of SNe Ia in passive hosts. The distinct different behaviors of $\text{pSNuM}_{\text{Ia, SF}}$ abd $\text{pSNuM}_{\text{Ia, P}}$ are consistent with the bimodal delayed-time distribution model of SNe Ia \citep{2006MNRAS.370..773M}.

\subsubsection{CC SN pSNuM}
\label{sec:pSNuM_CC}

\begin{figure}[htb!]
    \includegraphics[width = 0.98\linewidth]{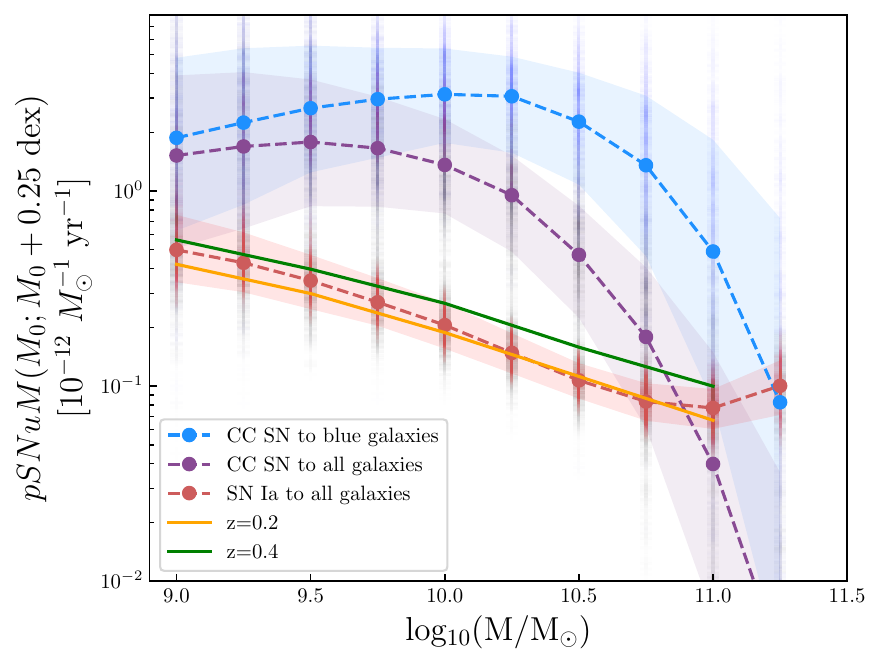}
    \caption{The pSNuMs of SN Ia ($z\sim0.22$) and CC SN ($z\sim0.08$) as functions of the lower bound of integral $M_{0}$ with an interval length of 0.25 dex. SN Ia is compared to all the field galaxies (red) \citep{2016MNRAS.459.2150W}, while CC SN is compared to all (purple) and blue population (blue) galaxies \citep{2012MNRAS.421..621B}, respectively. The orange and green curves show the SNuM predictions \citep{2021MNRAS.506.3330W}  at $z=0.2$  and $z=0.4$,  respectively. Each scatter mark an MC experiment result. Filled circles show the MC median values, lines represent the linear interpolation, and the shaded regions show the 16-84 percentile of the MC results.}
    \label{fig:pSNuM}
\end{figure}

\begin{figure}[htb!]
    \includegraphics[width = 0.98\linewidth]{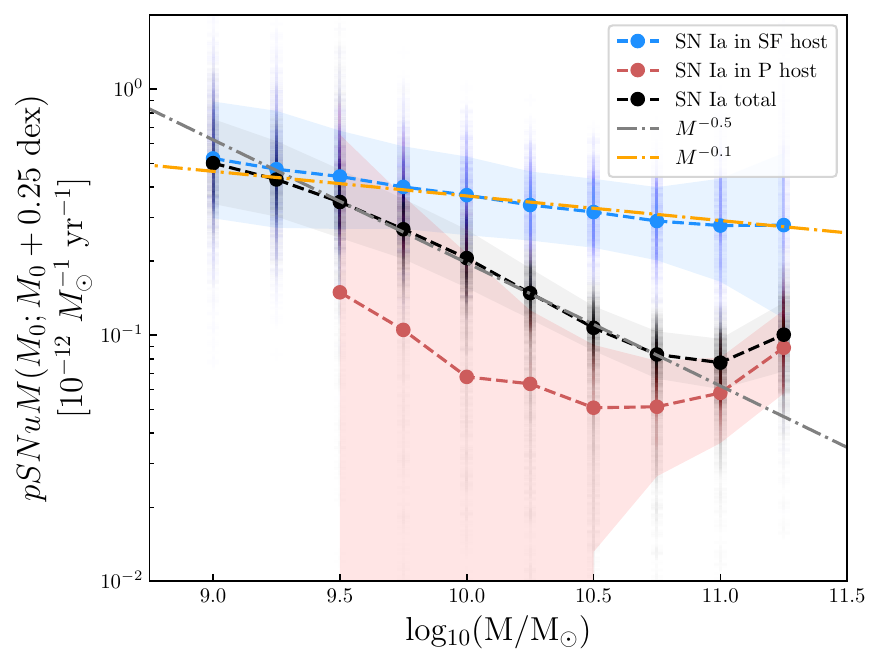}
    \caption{The SN Ia pSNuMs as functions of the lower bound of integral $M_{0}$ with an interval length of 0.25 dex. The total pSNuM$_{\text{Ia}}$ is compared to all the field galaxies (black)\citep{2016MNRAS.459.2150W}, the star-forming pSNuM$_{\text{Ia}}$ is compared to blue population field galaxies (blue) \citep{2012MNRAS.421..621B}, and the passive pSNuM$_{\text{Ia}}$  is compared to red population field galaxies (red) respectively. The dashed line shows a power-law with an index of $-0.5$ corresponding to \citet{2011MNRAS.412.1473L}. Each scatter mark an MC experiment result. Filled circles show the MC median values, lines represent the linear interpolation, and the shaded regions show the 16-84 percentile of the MC results.}
    \label{fig:pSNuMIa}
\end{figure}

The $\text{pSNuM}_{\text{CC}}$ remains nearly a constant rate below $10^{10.25} M_{\odot}$ and then rapidly decreases (when weighted by all the field galaxies, the purple curve in Figure \hyperref[fig:pSNuM]{8}). A similar result is obtained after removing the red population field galaxies (blue curve), but a rapid decreasing trend happens beyond $10^{10.75} M_{\odot}$. Since CC SNe are the fates of massive stars with a short delayed time since the post star formation, their rates can be roughly assumed to be proportional to the instant star formation rates. Therefore, the $\text{SNuM}_{\text{CC}}$ can trace the sSFR behavior. Within uncertainties, the behaviors of the derived  $\text{pSNuM}_{\text{CC}}$ are found consistent with the relation between sSFR and stellar mass of general field galaxies in the local universe\citep{2003MNRAS.341...54K, 2004MNRAS.351.1151B,2021MNRAS.500.2036K}, which also has a similar pattern with a knee at around  $10^{10.5} M_{\odot}$. We note that it is assumed that the redshift-depending CC rate is proportional to the universal SFR in Section \ref{subsec:EFF} to estimate the survey efficiency. However, this assumption does not contain stellar mass information. Therefore, it is independent of the $\text{pSNuM}_{\text{CC}}$-stellar mass relation found here.

\section{Conclusion}
\label{sec:conclusion}

In this study, we utilize one of the most complete and homogeneous SN samples by far for host galaxy statistics. We find the following conclusions:

\begin{itemize}
    \item Both luminosity functions and mass functions of SN host galaxies differ from those of normal field galaxies. The fitting parameters of luminosity (mass) functions indicate that the SN Ia host galaxies tend to be overall more luminous (massive), while CC SN host galaxies tend to be overall less luminous (massive).
    \item The luminous (massive) end of SN Ia host galaxies luminosity (mass) function is dominated by passive galaxies.On the other hand, star-forming galaxies dominate the faint-end slopes.
    \item The functions of CC SN host galaxies are dominated by star-forming galaxies.
    \item The SN Ia yield efficiency monotonically increases as stellar mass (luminosity) increases, while CC SN yield efficiency peaks at intermediate massive (luminous) galaxies.
    \item The $\text{pSNuM}_{\text{Ia, SF}}$ shows a weaker dependence on stellar mass in comparison with $\text{pSNuM}_{\text{Ia}}$, suggesting a population of SN Ia that follows the star-formation.
    \item The behavior of CC SN yield efficiency and $\text{pSNuM}_{\text{CC}}$ suggests that the CC rate is proportional to the star-formation rate.
\end{itemize}

Since the SDSS-II SN survey was optimized for SN Ia, this study is based on a uniform SN Ia sample set. On the other hand, CC SN samples are limited by CC SNe volume, and they are less uniform since CC SNe include various types of transients. In addition, because the host galaxies are in general more luminous than SNe, the sample efficiencies are overall limited by the SN and the faint galaxy.  

We have shown the SED fitting for SN host galaxies, from where the luminosity functions and the stellar mass functions are derived and are discussed with respect to SN host star-forming activities. It is notable that there are 'passive' hosts for CC SN based on our classification, which is in conflict with consensus. Although early-type hosts have been found for extremely rare cases \citep[$\lesssim 1\%$,][]{2022ApJ...927...10I}, the ratio between passive hosts to all in our SN samples (weighted by SN efficiency and volume) is about $8\%$ which is still much higher. Also, these galaxies do not form a large enough sample set for high-confidence statistics. Moreover, these galaxies are located near the boundary between star-forming and passive galaxies. It is likely that these galaxies are the results of the SED fitting uncertainty, as we lack filter bands that are sensitive to star-forming activities.

The faint-end slopes in our results show that the SNRs decrease toward the faint, low-mass end of galaxies. To understanding the SNR in faint host, it requires the knowledge of dwarf galaxy properties as well as the SN luminosity function in these faint hosts. However, the limited number of faint samples and large uncertainty of faint-end slopes render an incomprehensive conclusion here. One possible reason to account for the estimated low SNRs is that SNe are pre-identified as hostless for faint galaxies. But we also note that only a limited number of hosts are retrieved in HSC. Furthermore, the volume of faint end is limited by SDSS spectroscopy in this study, rendering SNe devoid of applicable redshift information from faint hosts. Without redshift information, the identification and classification of SNe will also become less accurate, leading to the loss of SN samples in faint galaxies.  Despite various causes of uncertainty, our results can serve as a preliminary reference for the SN host luminosity functions. Future studies on constraining the faint-end slope are encouraged to be answered with a larger sample set of SNe and SN hosts in the upcoming era of LSST \citep{2009arXiv0912.0201L}. \\

This study is supported by the Japan Society for the Promotion of Science (JSPS) KAKENHI Grants number 18H05223 (MD), 23H05432 (MD), 18K03696 (NS), 20H04730 (NS), 23H04494 (NS) and 23K03451 (NS). 

The Hyper Suprime-Cam (HSC) collaboration includes the astronomical communities of Japan and Taiwan, and Princeton University. The HSC instrumentation and software were developed by the National Astronomical Observatory of Japan (NAOJ), the Kavli Institute for the Physics and Mathematics of the Universe (Kavli IPMU), the University of Tokyo, the High Energy Accelerator Research Organization (KEK), the Academia Sinica Institute for Astronomy and Astrophysics in Taiwan (ASIAA), and Princeton University. Funding was contributed by the FIRST program from the Japanese Cabinet Office, the Ministry of Education, Culture, Sports, Science and Technology (MEXT), the Japan Society for the Promotion of Science (JSPS), Japan Science and Technology Agency (JST), the Toray Science Foundation, NAOJ, Kavli IPMU, KEK, ASIAA, and Princeton University. 

This paper makes use of software developed for Vera C. Rubin Observatory. We thank the Rubin Observatory for making their code available as free software at \url{http://pipelines.lsst.io}.

This paper is based on data collected at the Subaru Telescope and retrieved from the HSC data archive system, which is operated by the Subaru Telescope and Astronomy Data Center (ADC) at NAOJ. Data analysis was in part carried out with the cooperation of Center for Computational Astrophysics (CfCA), NAOJ. We are honored and grateful for the opportunity of observing the Universe from Maunakea, which has the cultural, historical and natural significance in Hawaii. 

Funding for the Sloan Digital Sky Survey V has been provided by the Alfred P. Sloan Foundation, the Heising-Simons Foundation, the National Science Foundation, and the Participating Institutions. SDSS acknowledges support and resources from the Center for High-Performance Computing at the University of Utah. SDSS telescopes are located at Apache Point Observatory, funded by the Astrophysical Research Consortium and operated by New Mexico State University, and at Las Campanas Observatory, operated by the Carnegie Institution for Science. The SDSS web site is \url{www.sdss.org}.

SDSS is managed by the Astrophysical Research Consortium for the Participating Institutions of the SDSS Collaboration, including Caltech, The Carnegie Institution for Science, Chilean National Time Allocation Committee (CNTAC) ratified researchers, The Flatiron Institute, the Gotham Participation Group, Harvard University, Heidelberg University, The Johns Hopkins University, L’Ecole polytechnique f\'{e}d\'{e}rale de Lausanne (EPFL), Leibniz-Institut f\"{u}r Astrophysik Potsdam (AIP), Max-Planck-Institut f\"{u}r Astronomie (MPIA Heidelberg), Max-Planck-Institut f\"{u}r Extraterrestrische Physik (MPE), Nanjing University, National Astronomical Observatories of China (NAOC), New Mexico State University, The Ohio State University, Pennsylvania State University, Smithsonian Astrophysical Observatory, Space Telescope Science Institute (STScI), the Stellar Astrophysics Participation Group, Universidad Nacional Aut\'{o}noma de M\'{e}xico, University of Arizona, University of Colorado Boulder, University of Illinois at Urbana-Champaign, University of Toronto, University of Utah, University of Virginia, Yale University, and Yunnan University.

\facilities{Subaru, Sloan}
\software{Astropy \citep{astropy:2013, astropy:2018, astropy:2022}, SExtractor \citep{1996A&AS..117..393B}, CIGALE \citep{2019A&A...622A.103B, 2022ApJ...927..192Y}, Numpy \citep{2020Natur.585..357H}, and Matplotlib \citep{4160265}}

\appendix
\section{Validation of SED fitting}
\label{sec: A1}
We acknowledge that SED fitting parameters are easily degenerate with optical data as the only input. To demonstrate the reliability of the parameters we are interested in, we present the validation for our SED modeling here.

To characterize the performance of the SED fitting, we compare our model SEDs with observed galaxy spectra. Although the majority of our sample galaxies are located at the redshift beyond the SDSS spectroscopic limit, spectra for 5 SN Ia hosts and 10 CC SN hosts in the local universe are retrieved. In order to understand how the lack of u-band impacts our fitting, we perform the SED modeling with the same parameters where high S/N \textit{ugriz} SDSS Petrosian photometry is available in SDSS. 

In Figure \hyperlink{fig: A1}{10}, four representative host examples are shown. In general, both HSC photometry and SDSS photometry provide similar model continuum components. These model SEDs and observed spectra are in reasonable agreement in the wavelength range from 4700 $\AA$ to 9000 $\AA$. Generally, an SED in this wavelength range is dominated by stellar emission and dust attenuation for these low-z SN hosts. Therefore, we conclude that our modeling provides robust estimations for $M_{\star}$ of the galaxies. 

\begin{figure}[h]
\label{fig: A1}
\begin{center}
\includegraphics[width=16cm]{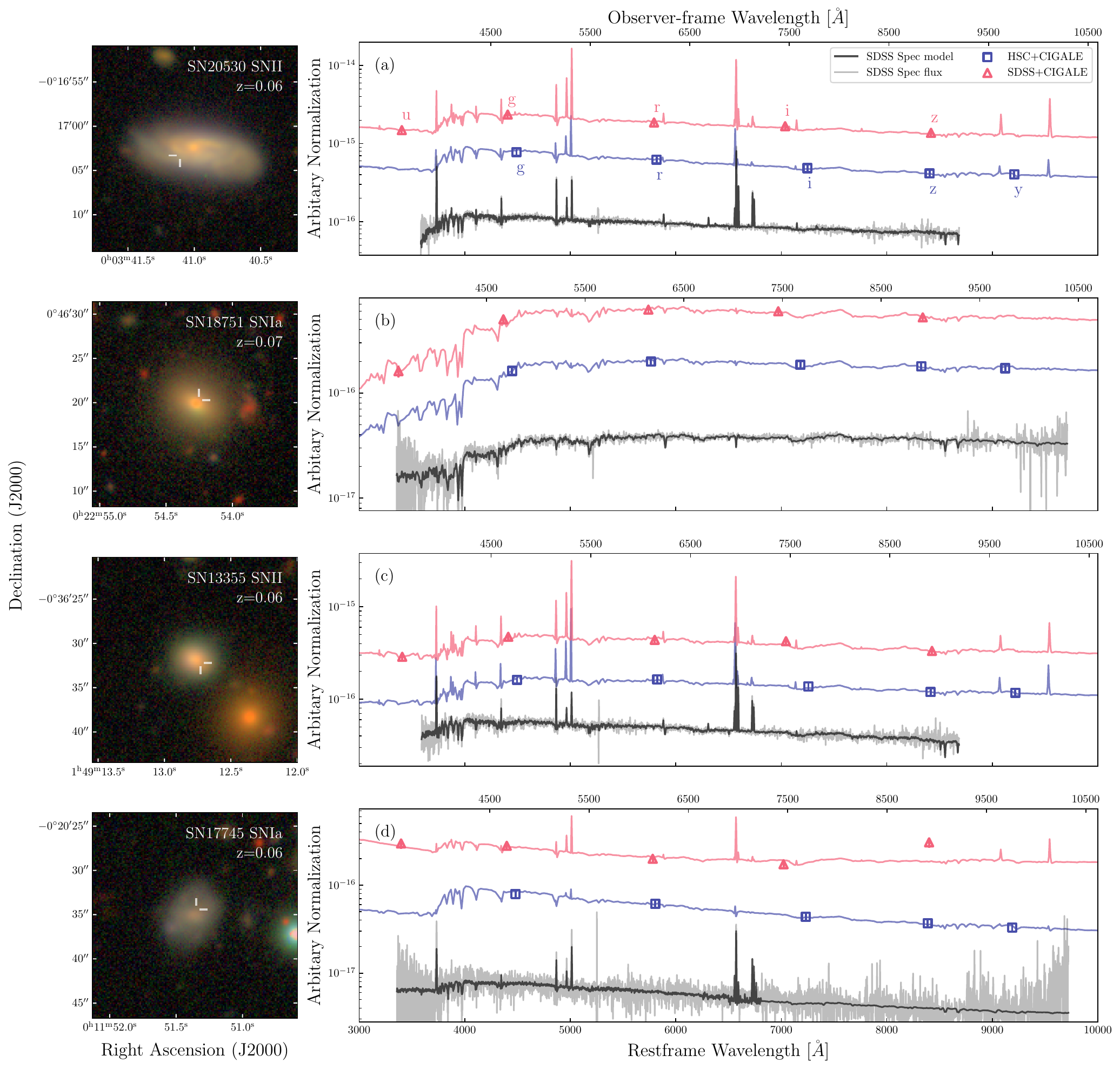} 
\end{center}
\caption{Examplary SED models and observed spectra for the nearby SN hosts. Blue squares and red triangles mark the HSC grizy and the SDSS ugriz observations as their SED fitting (lines encoded with the same colors) inputs. The SDSS pipeline model fits and the flux \citep{2000AJ....120.1579Y, 2011AJ....142...72E} of coadded spectra are shown in black and grey lines, respectively.}
\end{figure}

In the case of bright star-forming galaxies Figure \hyperlink{fig: A1}{10} (a) and (c) and a passive galaxy (b), both HSC and SDSS model SEDs tend to reproduce the observational spectra, especially at long wavelength. However, they tend to deviate from the spectrum model that fits at short wavelengths. The star-forming case (d) shows that the SDSS photometry is inconsistent with the spectrum, while the SED fitting with HSC photometry can reproduce it reasonably well. 

Inconsistency beyond Balmer-break is possibly contributed to the large uncertainty of SDSS spectra (Figure \hyperlink{fig: A1}{10} (b) and (d)), and to the difference in observational apertures. The latter one is able to explain the deficient UV component of BOSS spectrum in Figure \hyperlink{fig: A1}{10}(a), where spectroscopy targets only the bulge component of the galaxy (at the SN location). Whereas disk components are included in photometry, hence bluer models.

As it is suggested that the estimated sSFRs are biased when only using optical data \citep{2013ApJ...770..107C}, the sSFR is also loosely constrained in general and likely becomes unresolved when $\text{log}($sSFR$/[M_{\odot}yr^{-1}]) \leq -11$. Thus, it is difficult to acquire precise SFR. However, the classification of star-forming/passive bimodality consists well with the available observations.

Figure \hyperlink{fig: A2}{11} displays our identified passive CC SN host galaxies. It is clear that all these galaxies have relatively red colors, hence passive sSFRs. However, disk components are seen in  SN1127 and SN8569 hosts. And a merging host is identified for SN21596. Given the presence of dynamical structures as well as that they are late-type galaxies, these hosts are more likely to be star-forming. Apart from the clear examples of stellar formation activity mentioned above, it is difficult to determine whether the other two host galaxies are passive based on their morphology. The host galaxy of SN21567 resides in a galactic cluster and the crowded environment renders the SED fitting biased, as blending introduces undesirable uncertainty in photometry. Also, it is clear that the SN17477 host is an early-type galaxy, but the conclusion as to whether they are star-forming or passive remains elusive, as our SED fitting is not capable of resolving star-forming activities at the site of SN. Overall, the number of passive CC SN hosts (and the passive population in the derived luminosity function and mass function of CC SN hosts) is biased by the uncertainty of SED fitting. And no conclusive passive CC SN host is found in this work.

\begin{figure}[h]
\label{fig: A2}
\begin{center}
\includegraphics[width=16cm]{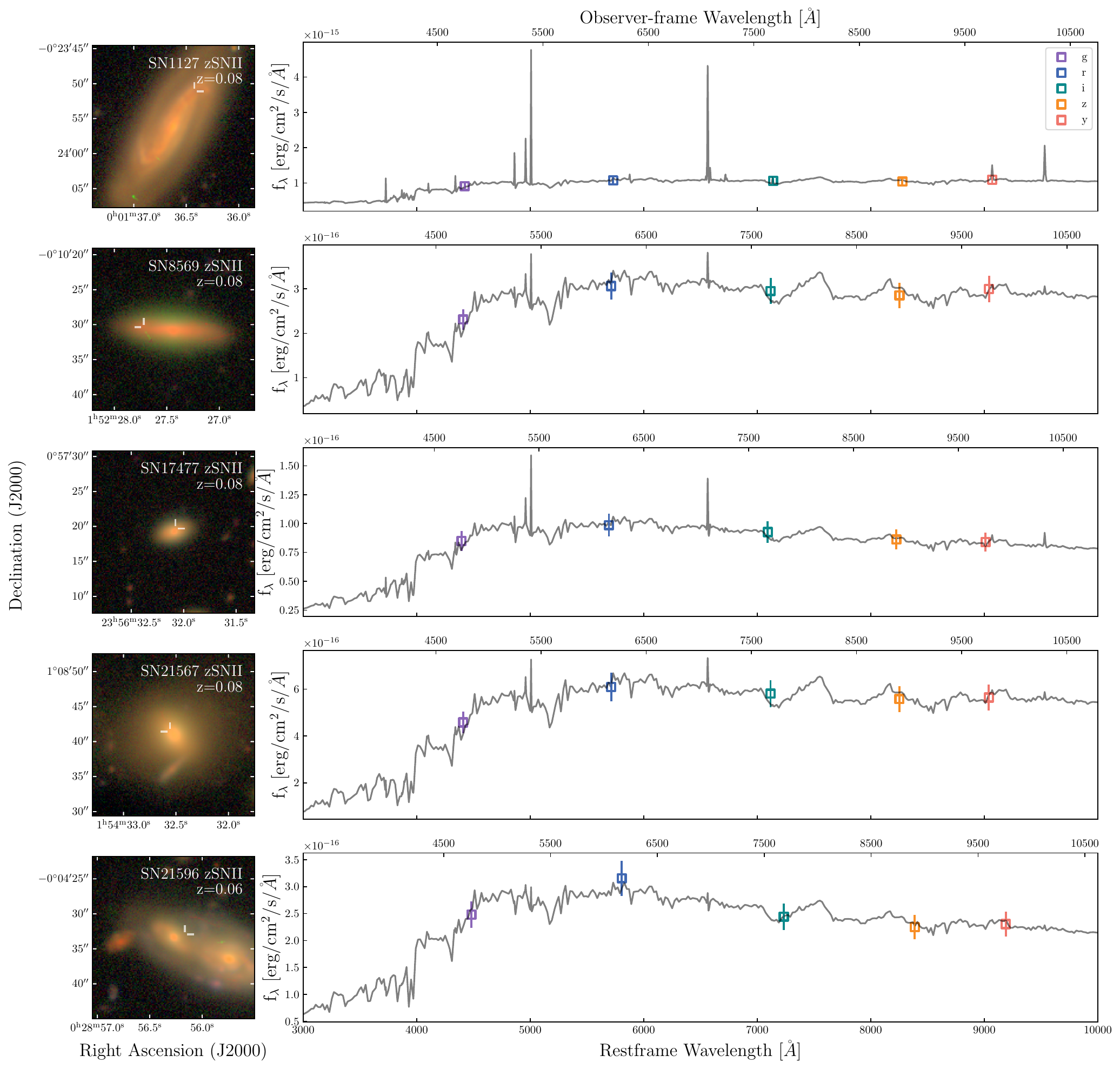} 
\end{center}
\caption{Passive CC SN hosts within sample selections.}
\end{figure}

\section{Mass Functions of SN Ia host galaxies}
\label{sec: A2}

The division of star-forming and passive SN Ia host galaxies allows us explore the SN Ia rate behavior in either kind of galaxies by their mass functions. However, the accurate fittings of mass functions are hindered by the limited entries of low-mass SN Ia host galaxies, making it difficult to find consistent faint-end slopes simultaneously. Given that it has relatively more detection in comparison with the passive group, we attempt to fit the mass function for the star-forming group with a Schechter function, and evaluate the passive group mass function as the difference between that of the total group and that of the star-forming group. We note that this method will overestimate the uncertainty of the passive group stellar mass function. In addition, we further assume that the faint end of total group mass function is dominating by star-forming galaxies. In practice, we fix the star-forming $\alpha$ to be $−0.606$, while $M_{\star}^{*}$ and $\phi_{\star}^{*}$ are free parameters. We find $M_{\star}^{*}=10.74\pm0.08\ M_{\odot}$ and $\phi_{\star}^{*}=(3.38\pm1.01)\times10^{-5}\ h^3\text{Mpc}^{-3}\text{dex}^{-1}\text{yr}^{-1}$. The decoupled mass functions are shown in Figure \hyperlink{fig: A3}{12}.

\begin{figure}[htb!]
\label{fig: A3}
\begin{center}
\includegraphics[width=8cm]{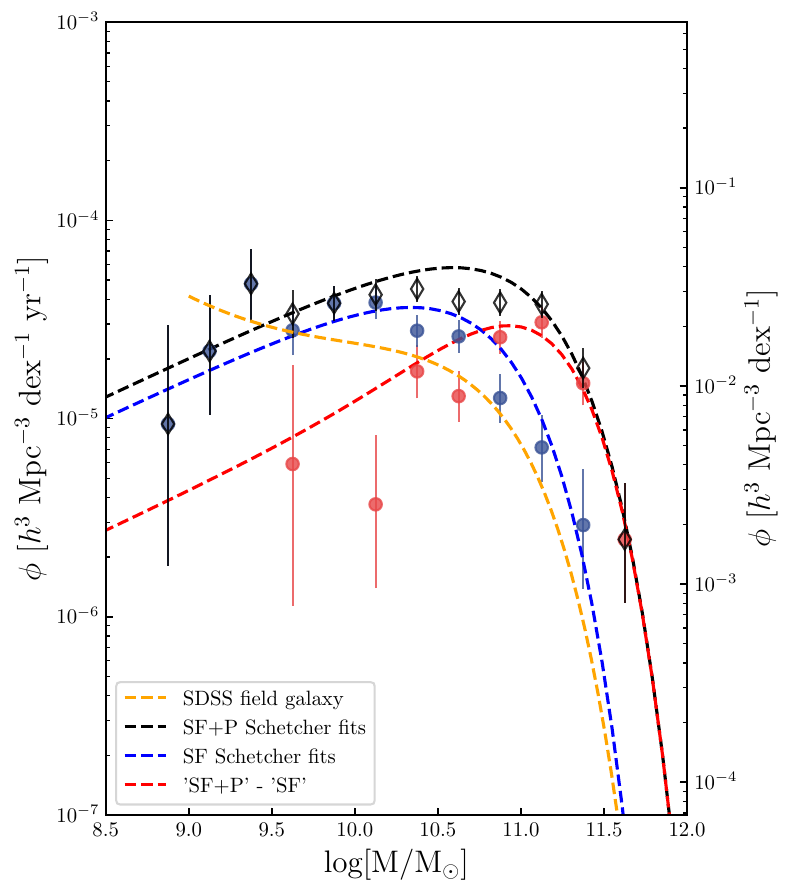} 
\end{center}
\caption{Decoupled SN Ia stellar mass function. The orange dashed curve show the mass function of field galaxy \cite{2016MNRAS.459.2150W}.} 
\end{figure}

\section{Supernova Host Galaxies Indentified in HSC-SSP Catalog}
Using HSC data and applying the redshift limit $z_{Ia} \leq 0.3$, $z_{CC} \leq 0.1$ as adopted in \hyperref[sec:LF]{$\S$4}, we also retrieved 30 hosts of spectroscopically confirmed SNe Ia that \citetalias{2018PASP..130f4002S} classifies as hostless SN Ia (Table \hyperref[tbl:hostless]{5}). These galaxies are in general fainter than $\sim$ 22 mag with one exceptional bright host of SN 2006fk ($r=19.34$ mag, $d_{DLR,r}=3.92$).
\begin{deluxetable}{lllllllllll}[ht!]
\tablecaption{Supernova Host Galaxies Identified in HSC-SSP Catalog}
\tablehead{
\colhead{SNID}&
\colhead{Name}&
\colhead{z$_{\text{CMB}}$}&
\colhead{Class}&
\colhead{$\alpha^{\text{SN}}_{J2000}$}&
\colhead{$\delta^{\text{SN}}_{J2000}$}&
\colhead{HSCID}&
\colhead{$\alpha^{\text{Host}}_{J2000}$}&
\colhead{$\delta^{\text{Host}}_{J2000}$}&
\colhead{r-CModel}&
\colhead{$d_{\text{DLR}}$} \\
\colhead{ }&
\colhead{ }&
\colhead{ }&
\colhead{ }&
\colhead{(deg)}&
\colhead{(deg)}&
\colhead{ }&
\colhead{(deg)}&
\colhead{(deg)}&
\colhead{(mag)}&
\colhead{ } 
}
\startdata
2943 & 2005go & 0.264 & SNIa & 17.7049 & 1.0079 & 41698450202434704 & 17.7052 & 1.0082 & 22.44$\pm$0.01 & 1.61 \\
3199 & 2005gs & 0.250 & SNIa & 333.2927 & 1.0505 & 42635238404285401 & 333.2925 & 1.0508 & 22.73$\pm$0.03 & 2.18 \\
6649 & 2005jd & 0.313 & SNIa & 34.2759 & 0.5348 & 41746682685183579 & 34.2758 & 0.5350 & 23.35$\pm$0.02 & 1.19 \\
6924 & 2005ja & 0.326 & SNIa & 358.9693 & 0.8770 & 42709726022098346 & 358.9693 & 0.8771 & 23.70$\pm$0.03 & 1.20 \\
6933 & 2005jc & 0.212 & SNIa & 11.3517 & 1.0756 & 41681274628236574 & 11.3514 & 1.0757 & 24.48$\pm$0.06 & 1.82 \\
7475 & 2005jn & 0.320 & SNIa & 4.7535 & -0.2815 & 40589871898717949 & 4.7534 & -0.2815 & 24.35$\pm$0.04 & 0.47 \\
12855 & 2006fk & 0.171 & SNIa & 330.2555 & 0.7162 & 42626571160268723 & 330.2552 & 0.7159 & 19.34$\pm$0.00 & 3.92 \\
13038 & 2006gn & 0.103 & SNIa & 347.8268 & 0.5046 & 42679480862378700 & 347.8267 & 0.5046 & 24.54$\pm$0.06 & 1.11 \\
13518 & 2006mn & 0.298 & SNIa & 16.9516 & -0.1098 & 40624785687865407 & 16.9517 & -0.1098 & 23.65$\pm$0.03 & 0.45 \\
13641 & 2006hf & 0.209 & SNIa & 345.2187 & -0.9811 & 41601684589269482 & 345.2184 & -0.9813 & 23.25$\pm$0.03 & 3.38 \\
15002 & 2006ko & 0.361 & SNIa? & 22.2498 & 0.7699 & 41711640046998805 & 22.2499 & 0.7698 & 22.13$\pm$0.01 & 1.08 \\
15674 & 2006nu & 0.197 & SNIa & 340.8292 & 0.2629 & 42657207162017702 & 340.8292 & 0.2629 & 23.70$\pm$0.02 & 0.55 \\
15734 & 2006ng & 0.383 & zSNIa & 5.5303 & -0.0956 & 40594823996010442 & 5.5302 & -0.0957 & 24.19$\pm$0.05 & 1.40 \\
16000 & 2006nj & 0.389 & SNIa & 21.1176 & 0.0744 & 41706949942729816 & 21.1174 & 0.0743 & 22.66$\pm$0.02 & 1.38 \\
16165 & 2006nw & 0.156 & SNIa & 30.7331 & -0.5339 & 40669582196761792 & 30.7331 & -0.5339 & 23.72$\pm$0.02 & 0.12 \\
16213 & 2006oi & 0.249 & SNIa & 8.9711 & 0.2584 & 41672044743521316 & 8.9711 & 0.2585 & 23.10$\pm$0.02 & 0.82 \\
17223 & 2007jj & 0.234 & SNIa & 26.3614 & -0.0194 & 40656400942123762 & 26.3616 & -0.0194 & 23.51$\pm$0.03 & 1.18 \\
17391 & 2007jo & 0.189 & SNIa & 347.5524 & -0.9314 & 41610892999153273 & 347.5525 & -0.9313 & 23.55$\pm$0.05 & 0.96 \\
17825 & 2007je & 0.160 & SNIa & 32.9471 & -0.9124 & 40673421897521423 & 32.9469 & -0.9123 & 24.15$\pm$0.05 & 1.54 \\
18451 & 2007mt & 0.387 & SNIa? & 26.4858 & -0.2176 & 40656259208203172 & 26.4858 & -0.2179 & 22.70$\pm$0.01 & 1.63 \\
18945 & 2007nd & 0.269 & SNIa & 10.0784 & -1.0374 & 40607996660697189 & 10.0784 & -1.0374 & 23.56$\pm$0.03 & 0.32 \\
19128 & 2007lw & 0.285 & SNIa & 354.2038 & -0.7822 & 41627939724350117 & 354.2037 & -0.7823 & 24.13$\pm$0.04 & 1.08 \\
19282 & 2007mk & 0.176 & SNIa & 359.0722 & -0.5059 & 41640863280946029 & 359.0723 & -0.5058 & 23.71$\pm$0.04 & 1.01 \\
20186 & 2007pj & 0.352 & SNIa & 357.2946 & 0.7979 & 42705461119577449 & 357.2945 & 0.7974 & 22.86$\pm$0.02 & 4.02 \\
20345 & 2007qp & 0.264 & SNIa & 10.7018 & 0.3797 & 41676172207088916 & 10.7021 & 0.3798 & 23.63$\pm$0.03 & 3.13 \\
20934 & 2007rm & 0.301 & SNIa & 35.4379 & 0.8645 & 41751364199515663 & 35.4379 & 0.8644 & 23.79$\pm$0.04 & 0.29 \\
20978 & 2007rl & 0.323 & SNIa & 35.3878 & -0.3749 & 40682643192305770 & 35.3878 & -0.3749 & 22.79$\pm$0.01 & 0.35 \\
21033 & 2007qy & 0.228 & SNIa & 28.8137 & 0.6431 & 41728678182264991 & 28.8136 & 0.6430 & 23.49$\pm$0.02 & 0.81 \\
21422 & 2007rq & 0.266 & SNIa & 13.3838 & 0.9003 & 41685256062897311 & 13.3838 & 0.9006 & 22.30$\pm$0.01 & 1.99 \\
21810 & 2007se & 0.174 & SNIa & 333.1546 & 0.7967 & 42635367253311407 & 333.1547 & 0.7970 & 21.86$\pm$0.01 & 1.29 \\
\enddata
\end{deluxetable}

\label{tbl:hostless}


\clearpage
\bibliography{references}{}
\bibliographystyle{aasjournal}
\end{document}